\documentclass[useAMS,usenatbib]{mn2e}
\usepackage{times}
\usepackage{graphicx, latexsym, amssymb, amscd, psfrag}
\usepackage{epsfig}
\usepackage{color}
\title[FR I radio galaxy 3C 270]{New insights into the evolution of the FR I radio galaxy 3C 270 (NGC~4261) from VLA and GMRT radio observations}

\author[Konstantinos Kolokythas et al.]
{Konstantinos Kolokythas$^{1}$\thanks{e-mail: kkolok@star.sr.bham.ac.uk}, Ewan O'Sullivan$^{2,1}$, Simona Giacintucci$^{3,4}$, 
\newauthor{Somak Raychaudhury$^{5,1}$, C.~H.~Ishwara-Chandra$^{6}$, Diana M. Worrall$^{7,2}$ and } 
\newauthor{Mark Birkinshaw$^{7,2}$}\\ \\
$^{1}$School of Physics and Astronomy, University of Birmingham,
  Birmingham B15~2TT, UK\\
$^{2}$Harvard-Smithsonian Center for Astrophysics, 60 Garden Street,
Cambridge, MA 02138, USA\\
$^{3}$Department of Astronomy, University of Maryland, College Park, MD, 20742-2421, USA\\
$^{4}$Joint Space-Science Institute, University of Maryland, College Park, MD, 20742-2421, USA\\
$^{5}$Department of Physics, Presidency University, 86/1 College Street, 700 073 Kolkata, India\\
$^{6}$National Centre for Radio Astrophysics, TIFR, Post Bag No. 3, Ganeshkhind, Pune 411 007, India\\
$^{7}$HH Wills Physics Laboratory, University of Bristol, Tyndall Avenue, Bristol BS8 1TL, UK }

\begin{document}

\date{Accepted ....; Received ....; in original form ....}

\pagerange{\pageref{firstpage}--\pageref{lastpage}} \pubyear{2014}
\maketitle

\label{firstpage}
\begin{abstract}
We present \textit{Giant Metrewave Radio Telescope} (GMRT) 240~MHz observations of the nearby luminous FR~I radio source 3C~270, in the group-central elliptical NGC 4261. Combining these data with reprocessed \textit{Very Large Array} (VLA) 1.55 and 4.8~GHz observations, we produce spectral index maps that reveal a constant spectral index along the jets and a gradual steepening from the ends of the jets through the lobes towards the nucleus.  A Jaffe \& Perola (JP) model fitted to the integrated spectrum of the source gives an asymptotic low-frequency index of $\alpha_{inj}=0.53_{-0.02}^{+0.01}$, while JP models fitted to the observed spectral index trend along the lobes allow us to estimate radiative ages of $\sim29$~Myr and $\sim37$~Myr for the west and east lobes respectively. Our age estimates are a factor of two lower than the 75-Myr upper limit derived from X-ray data \citep{ewan4261}. We find unlikely the scenario of an early supersonic phase in which the lobe expanded into the ISM at approximately Mach 6 (3500 km s$^{-1}$), and suggest that either the source underwent multiple AGN outbursts with possible large changes in jet power, or possibly that the source age that we find is due to a backflow that transports young electrons from the jet tips through the lobes toward the nucleus relatively quickly. We calculate that in the lobes the energy ratio of non-radiating to radiating particles is $\sim4-24$ indicating significant gas entrainment. If the lobes are in pressure balance with their surroundings, the total energy required to heat the entrained material is $10^{58}$~erg, $\sim$40\% of the total enthalpy of the lobes.
\end{abstract}


 
\section{Introduction}
\label{sec:intro}
 
The study of radio galaxies has become increasingly important as they appear to be natural laboratories uniquely suited to the study of one of the most energetic phenomena in the Universe: relativistic jets arising from the associated active galactic nuclei (AGN). The reason this phenomenon is so important is that radio-loud activity of the central AGN is now considered to be the most likely source of the heating mechanism injecting energy into the hot gaseous halos of groups and clusters of galaxies, thereby balancing the effects of radiative cooling (\citealt{Fabian12}; \citealt{McNamara07}; \citealt{Peterson06}). 

The evolution of radio galaxies is commonly accepted to pass through three phases \citep[as described in, e.g.,][]{Kraft07}. In the beginning, the inflation of the lobes surrounding the jets is highly supersonic as the lobes appear to be significantly over-pressured in relation to the ambient medium. After this early, in most sources short-lived, supersonic phase, the expansion of the lobes continues and their internal pressure drops, approaching equilibrium with the ambient gas. In the final phase jets shut down and the lobes eventually become unobservable due to energy losses. Examples of the complex interactions between radio lobes and a surrounding intra-cluster or intra-group medium (IGM) are relatively common in the local universe. Radio galaxies in groups and clusters are observed to drive shocks into the IGM (\citealt{Nulsen05}; \citealt{Fabian06}; \citealt{Forman07}; \citealt{David11}) which in some cases are detectable as X-ray surface brightness discontinuities (e.g. in 3C~310, \citealt{Kraft12} and PKS~B2152-699, \citealt{Worrall12}), and to inflate cavities (e.g. \citealt{Birzan08}; \citealt{OSullivan11}; \citealt{Fabian12}).

Present knowledge of the physical processes acting behind this phenomenology is still rather poor. In order to understand how AGN feedback works, the combination of high-quality X-ray data and high-sensitivity radio observations is essential. X-ray observations provide information on the physical properties of the environment surrounding the radio source, most notably the pressure of the IGM that the radio lobes are expanding into, and allow dynamical estimates of the expansion timescale and AGN power output. Multi-frequency radio data provide information on the cycle of jet activity, including independent estimates of the source expansion timescale and indications of the physical mechanisms of energy injection \citep{Giacintucci08,Giacintucci12}.

With this aim we have chosen to examine NGC 4261, a nearby galaxy which has an active galactic nucleus that contains a supermassive black hole of (4.9$\pm$1.0)$\times$10$^{8}M_{\odot}$ \citep{Ferrarese96} and hosts the low-power FR I radio source 3C 270. It is the brightest elliptical in a poor group \citep{Forbes06} projected onto the outskirts of the nearby Virgo cluster. Its proximity means that its features can be observed with good sensitivity and linear resolution, allowing detailed study of the energetics of lobe inflation. 

3C 270 is composed of symmetric narrow kiloparsec-scale twin jets
  \citep{Bir85} which previous modelling studies indicate lie close
    to the plane of the sky (63$^{\circ}-76^{\circ}$; \citealt{Piner01},
    \citealt{Laing14}), and which flare and broaden before extending tens
    of kiloparsecs into the intragroup medium (IGM) \citep{Worrall10}. The
  jets emanate from a compact radio core coincident with the optical
  nucleus of the galaxy \citep{JonesWehrle97}, where optical spectroscopy
  has revealed a low-ionization nuclear emission-line region
  \citep{HoFilippenkoSargent97}. In addition, a nearly edge-on nuclear disk
  (100-pc-scale) rich in dust, molecular and atomic gas \citep{Jaffe93},
  lies orthogonal to the radio jets \citep{Tremblay09}. \citet{Laing13}
  examined the 1.4-4.8~GHz spectral index of the base of the jets and found
  values of $\sim$0.65 at $\sim$10$^{''}$ from the nucleus, jumping to 0.72
  at $\sim$14$^{''}$ and then declining to $\sim$0.6-0.65 at 50$^{''}$. The spectral index is defined as $S_\nu \propto \nu^{-\alpha}$, where $S_\nu$ is the flux density at the frequency $\nu$.

In the X-ray band, ROSAT images of the NGC~4261 group show diffuse thermal emission and a non-thermal nuclear component at low energies \citep{WorrallBirkinshaw94}. Chandra observations reveal X-ray jets in the inner few kiloparsecs of the radio jets \citep{Gliozzi03,Zezas05} and suggest a synchrotron origin for the jet X-ray emission.

More recent analysis of the \textit{Chandra} and \textit{XMM-Newton} data in \citet{ewan4261} shows much structure in the gas in which the galaxy is embedded. The AGN jets have inflated two lobes, producing cavities in the IGM and building up rims of compressed hot gas which almost entirely enclose the lobes. The cavity rims consist of material that has been swept up and compressed by the expansion of the lobes, weakly heated by compression with no evidence of shock heating  \citep{ewan4261}. NGC 4261 hosts a small cool core with radius $\sim10$ kpc and a temperature of $\sim0.6$ keV, with the IGM temperature in the group being $\sim1.6$ keV \citep{Humphrey09}. The structure of the core has been disturbed by the expansion of the radio lobes and is not spherically symmetric \citep{ewan4261}, with wedge-shaped surface-brightness decrements that suggest the jets have driven out the IGM (or galactic gas) from conical regions around their first few kiloparsecs \citep{Worrall10}.  

Evidence of anisotropy in the globular cluster distribution has been seen (\citealt{Giordano05}, \citealt{Bonfini12}, \citealt{DAbrusco13}) while deep optical imaging has revealed a weak tidal tail to the north-west and a tidal fan extending southeast from the galaxy \citep{Tal09}. These disturbed features suggest that NGC~4261 went through merging or tidal interactions with another galaxy within the past 1-2 Gyr.

In this paper we present a detailed low-radio-frequency spectral study of
the galaxy NGC 4261. Previous studies have left the question of the
radiative age of the source unclear and with our study will try to resolve
this problem. Our work is based on archival data from the Giant Metrewave
Radio Telescope (GMRT) at 610 MHz and 240 MHz and the Very Large Array
(VLA) at 1.55 GHz and 4.8 GHz. The primary goals of this work are to
determine the radiative age of 3C 270 and to understand the energetics and
dynamics of the interaction between the relativistic plasma of the radio
lobes and the X-ray emitting gas following \citet{ewan4261}. The paper is
organised as follows. In Sections~\ref{sec:GMRT} and \ref{sec:VLA} we
describe the GMRT and VLA archival data used and the data reduction. The
radio images are presented in Section~\ref{sec:radims}. In
Section~\ref{sec:analysis} we derive the physical properties of the
  radio source. In Section~\ref{sec:discuss} we discuss our
  results and consider the development of the radio source in the context
  of its environment. The summary and conclusions are finally presented in
Section~\ref{sec:conc}.

We adopt a distance to NGC 4261 of 31.3 Mpc, which gives an angular scale of $ 1^{\prime\prime} = 0.151$ kpc. This is consistent with the distance adopted by \citet{Worrall10} and negligibly different from the distance adopted by \cite{ewan4261} who used an angular scale of $ 1^{\prime\prime} = 0.153$ kpc for their X-ray analysis. When correcting for redshift we adopt a value of z=0.007378 \citep{Cappellari11}. The radio spectral index $\alpha$ is defined according to $S_\nu \propto \nu^{-\alpha}$, where $S_\nu$ is the flux density at the frequency $\nu$.

\begin{table*}
\begin{center}
\caption{\label{tab:gmrtobs}Summary of GMRT radio observations}
\begin{tabular}{lcccccc}
\hline
 Project Code &Observation date &Frequency&Bandwidth&Integration time & HPBW\footnote{Using uniform weighting}, PA& rms \\
  &        & (MHz)   & (MHz)   & (min)& (full array, $''\times'',{}^{\circ}$)& (mJy beam$^{-1}$) \\
\hline
 15KMH01&2009 Feb 15&610&32& 270&7.32$\times$4.77, 76.62& 1\\ 
 15KMH01&2009 Feb 15&240&16& 270&15.30$\times$11.00, 67.17& 1.6  \\ 
\hline
\end{tabular}
\end{center}
\end{table*}

\section{GMRT observations and data reduction}
\label{sec:GMRT}

\begin{figure}
\includegraphics[trim=0.3cm 0cm 0cm 0cm, clip=true,width=\columnwidth]{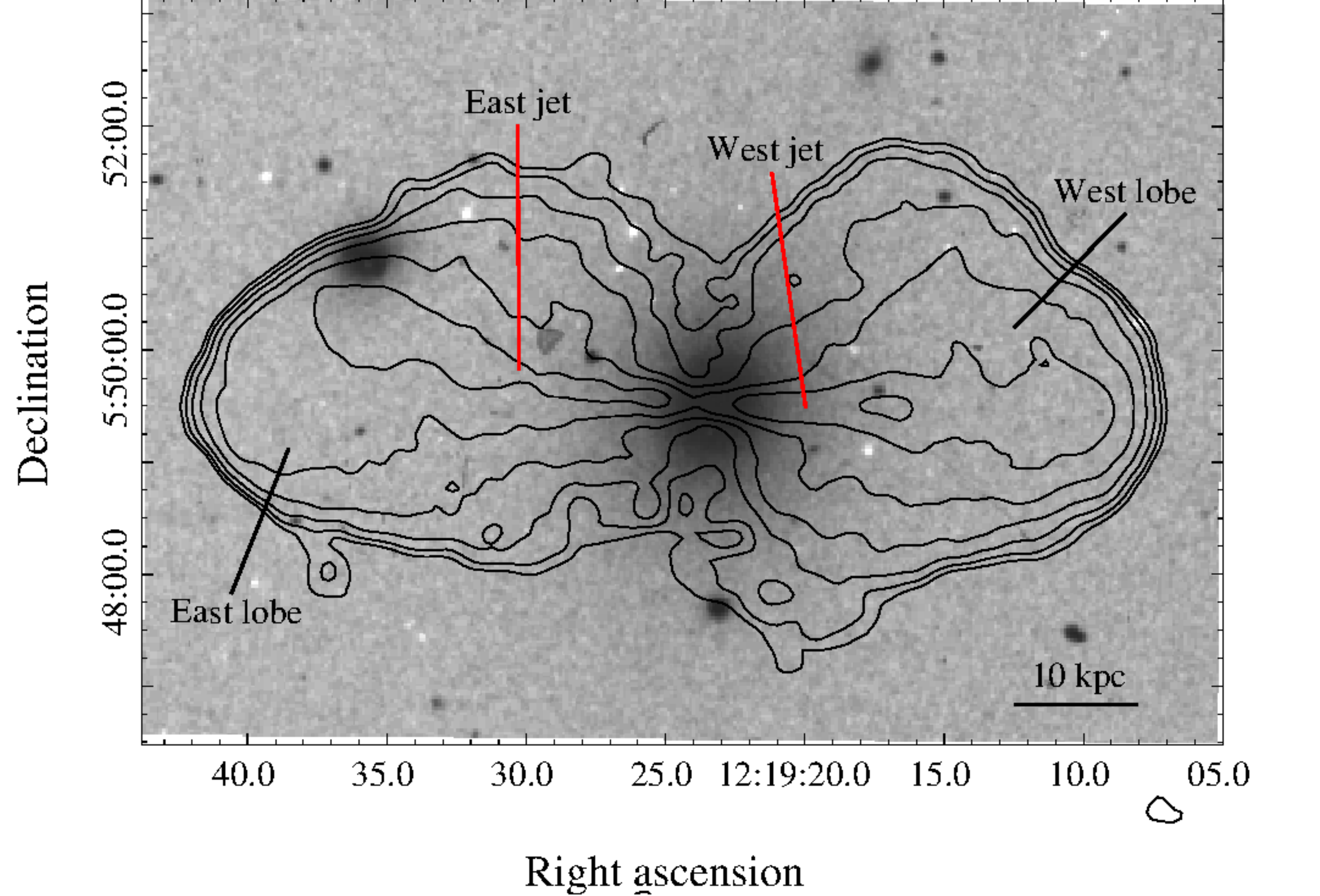}
\caption{\label{fig:labels}GMRT 240 MHz radio contours of 3C 270, overlaid on the $R$ band optical image from the Digitized Sky Survey (DSS). The contour levels (black) start at 4$\sigma$ and rise by factors of 2. The 1$\sigma$ level in the radio image is $\sim$ 1.6~mJy~beam$^{-1}$.  Labels indicate the individual components of the radio source. Restoring beam: $15.30\times11.00$, P.A.= 67.17$^{\circ}$.}
\end{figure}

NGC 4261 was observed using the GMRT at 240 MHz and 610 MHz with total integration time of 4.5 hours in each band. At 610 MHz the data were recorded using the lower and upper side bands for an observing bandwidth of 16 MHz each. For the 240 MHz data only one correlator was used to record the data in a total observing bandwidth of 16 MHz where only half of the data were useful. Data for both frequencies were collected with 256 channels at a spectral resolution of 125 kHz per channel. The data were analysed and reduced using the NRAO Astronomical Image Processing System (AIPS) package. We summarize the details of the observations in Table~\ref{tab:gmrtobs}, where we report the observing date, frequency, total observing bandwidth, total time on source, half-power beamwidth (HPBW) of the full array and rms level (1 $\sigma$) at full resolution. 

The data were edited to remove bad antennas and other obvious
  defects. Initially the phase and flux calibrators were checked and the
  data with extreme phase differences were removed. This procedure was also
applied to our source. The data were calibrated and the flux
density scale was defined using the amplitude calibrators 3C~286 and
3C~147. The task SETJY was used to set the flux density scale to
  ``best VLA values (2010)''. The flux densities set for 3C~286 are 25.84$\pm$0.7 Jy for 240 MHz, 14.54$\pm$0.2 Jy for 1.5~GHz and 7.46$\pm$0.05 Jy for 4.8~GHz, while for 3C~147 the flux density at 240 MHz is 48.65$\pm$1.46 Jy. The calibrator flux densities agree with \citet{Perley13} to within 1\% and the \citet{ScaifeHeald12} scale to within 3\% except for 3C~147 at 240MHz, where the two
scales disagree (60.95$\pm$2.26 Jy for \citealt{ScaifeHeald12} at 240 MHz). The point source 1330+251 was used as phase
calibrator for both 610 and 240 MHz. The bandpass was calibrated using observations of
3C~286, with channels that presented amplitude `spikes' removed. At 610 MHz about 25\% of the data
were edited out. In order to increase signal-to-noise per channel during RFI removal while limiting
the effects of bandwidth smearing, the channels were averaged into
6 of $\sim$2 MHz each at 610 MHz, and 4 of $\sim$1.5 MHz each at 240 MHz.


Due to the large field of view of the GMRT, the field was split into multiple facets for imaging. 56 facets were created at 610 MHz (with cellsize 1.5$''$) and 85 facets were created at 240 MHz (with cellsize 3$''$) so that in each facet the imaging field still approximates a plane surface. When recombined for our final image these give a field of view of $\sim 1.2^{\circ}\times1.2^{\circ}$ at 610 MHz and $\sim 3^{\circ}\times3^{\circ}$ at 240 MHz. After editing again the averaged data, repeated cycles of deconvolution were performed along with careful phase-only self-calibration in order to image the data. Most of the remaining noise in our final images arises from confusion and calibration uncertainties. At the lowest frequencies the phase errors are expected to originate from rapidly varying ionospheric delays.

The final images were corrected for the primary beam pattern of the GMRT using the task PBCOR in AIPS. The rms noise level (1$\sigma$) achieved in the final full resolution images is 1 mJy at 610 MHz, and 1.6 mJy at 240 MHz (see Table~\ref{tab:gmrtobs}).

\citet{Chandra04} estimate the typical uncertainty on GMRT flux
  measurements to be $\sim$5\% at 610 MHz and $\sim$8\% at 240 MHz,
  including contributions from the uncertainty in the measurement of
  calibrator flux and uncertainties introduced by the telescope hardware.
  These are the levels of uncertainty which we quote in
  Table~\ref{tab:radprop}. In addition, they quote an elevation dependent
  error on the scale of 2-4$\%$, arising from changes in telescope response
  between the calibrator and source positions. These errors dominate over
  the systematic errors in the flux density scale.

Examination of the 610-MHz maps revealed ghost images containing up
  to $\sim$10\% of the flux of the main image to the N and S of the true
  radio structure, presumably because of residual delay and amplitude
  errors, though no attempt to improve these corrections was successful.
  The 610-MHz image is generally similar to images with comparable angular
  resolution from the VLA (Section~\ref{sec:VLA}), but since the ghost
  images introduce false structure, we choose to use the 610 MHz data only
  for an estimate of the total flux density, and not for spectral analysis
  of regions within the source

We imaged the phase and flux calibrators to check the integrity of the calibrations. While these were as expected, the flux densities derived for 8 background sources in the 3C~270 field were found to be systematically low by about 20\% at 240 and 610 MHz as compared with expectations based on literature measurements. At 240 MHz all source structures appeared normal, and it was considered appropriate to increase all flux densities, including those for 3C~270, by 20\%. The same procedure was applied for the total flux density of 3C~270 at 610 MHz. We note that this discrepancy cannot be resolved by using the higher 240~MHz calibrator flux for 3C~147 suggested by the \citet{ScaifeHeald12} scale, since this only affects one of the two calibrators, and only at one frequency.

\section{VLA observations and data reduction}
\label{sec:VLA} 

From the VLA public archive we acquired and reanalysed the radio
observations listed in Table~\ref{tab:vlaobs}. The data were calibrated and
imaged in AIPS. In order to achieve the best possible noise level and $uv$
coverage for the 1.55 GHz and 4.86 GHz VLA archival data, several datasets
were combined for mapping.  Individual observations were analysed
  separately until the stage where the source and calibrator $uv$ data were
  separated. The SPLIT files were combined using DBCON and the analysis
proceeded with phase-only self-calibration and imaging. For the
  1.55 GHz data, as can be seen in Table~\ref{tab:vlaobs}, four different
  observations were combined together (AP0077, AV0088, AH0343 A and AH0343
  B) in order to produce the final image with a total integration time of
  61 minutes and an rms noise level of 0.8 mJy (Array B\&C in
  Table~\ref{tab:vlares}). At 4.86 GHz only the D-array observation was
  used and analysed as it was sufficiently long and has a resolution
  comparable to the combined 1.55 GHz image. The properties of the
resulting maps are given in Table~\ref{tab:vlares}.

\begin{table*}
\begin{center}
\caption{\label{tab:vlaobs}Summary of archival VLA observations}
\begin{tabular}{lccccc}
\hline
 Project  & Observation date & Array & Frequency & Bandwidth      & Integration time \\
          &                  &       &  (GHz)    &    (MHz)       &  (min) \\
\hline
 AP0077   & 1984 Apr 23 &   C   &    1.55    &     2$\times$50        &     25              \\ 
 AV0088   & 1984 Jun 05 &   C   &    1.49    &     2$\times$50        &     16              \\ 
 AH0343 A & 1989 Mar 15 &   B   &    1.57    &     2$\times$50        &     10                \\
 AH0343 B & 1989 Jul 28 &  BnC  &    1.57  &     2$\times$50       &     10              \\
 AL0693 B & 2007 Mar 18 &   D   &    4.86    &     2$\times$50        &     81              \\ 
\hline
\end{tabular}
\end{center}
\end{table*}

The task LGEOM was used to rotate the data where needed, to match the VLA and GMRT observations for spectral index mapping. Finally the images were corrected for the primary beam response using the task PBCOR. Both images have spatial resolution similar to that of the GMRT 240 MHz image. We again adopted the ``best VLA values (2010)'' flux density scale.

\begin{table*}
\begin{center}
\caption{\label{tab:vlares}Summary of reduced VLA images}
\begin{tabular}{lcccc}
\hline 
Array & Frequency & Integration time &          HPBW, PA         & rms \\
 
      &  (GHz)    &  (min)           & ($''\times'',{}^{\circ}$) & (mJy beam$^{-1}$) \\
\hline
B\&C  &   1.55    &     61           & 19.0$\times$11.2, -55.35  & 0.8 \\ 
D     &   4.86    &     81           & 15.8$\times$10.7, -57.44  & 0.4 \\ 
\hline
\end{tabular}
\end{center}
\end{table*}

\begin{table*}
\begin{center}
\caption{\label{tab:radprop}Radio properties of NGC~4261}
\begin{tabular}{lccccccc}
\hline
 Components& $S_{240}$ MHz  & $S_{1.55}$ GHz & $S_{4.8}$ GHz & $\alpha^{240 \rm{MHz}}_{1.55 \rm{GHz}}$ & $\alpha^{1.55 \rm{GHz}}_{4.8 \rm{GHz}}$ & projected size\\
 
           &     (Jy)       &       (Jy)     &         (Jy)  &               &               & (kpc)\\
\hline
 Total ....& 48.50$\pm$3.21 & 18.30$\pm$0.92 & 8.30$\pm$0.42 & 0.41$\pm$0.08 & 1.01$\pm$0.07 & $80\times40$ \\
 Core .....&  0.48$\pm$0.04 &  0.36$\pm$0.02 & 0.37$\pm$0.02 & 0.16$\pm$0.10 & 0.03$\pm$0.08 & - \\
 West jet..& 10.60$\pm$0.85 &  3.80$\pm$0.19 & 1.38$\pm$0.20 & 0.47$\pm$0.09 & 0.90$\pm$0.15 & $37\times7$ \\
 East jet..&  8.00$\pm$0.64 &  2.80$\pm$0.14 & 1.13$\pm$0.06 & 0.56$\pm$0.09 & 0.80$\pm$0.07 & $43\times8$ \\
\hline
\end{tabular}
\end{center}
\end{table*}

\begin{figure}
\includegraphics[trim=11.2cm 2cm 15.3cm 0cm, clip=true,width=\columnwidth]{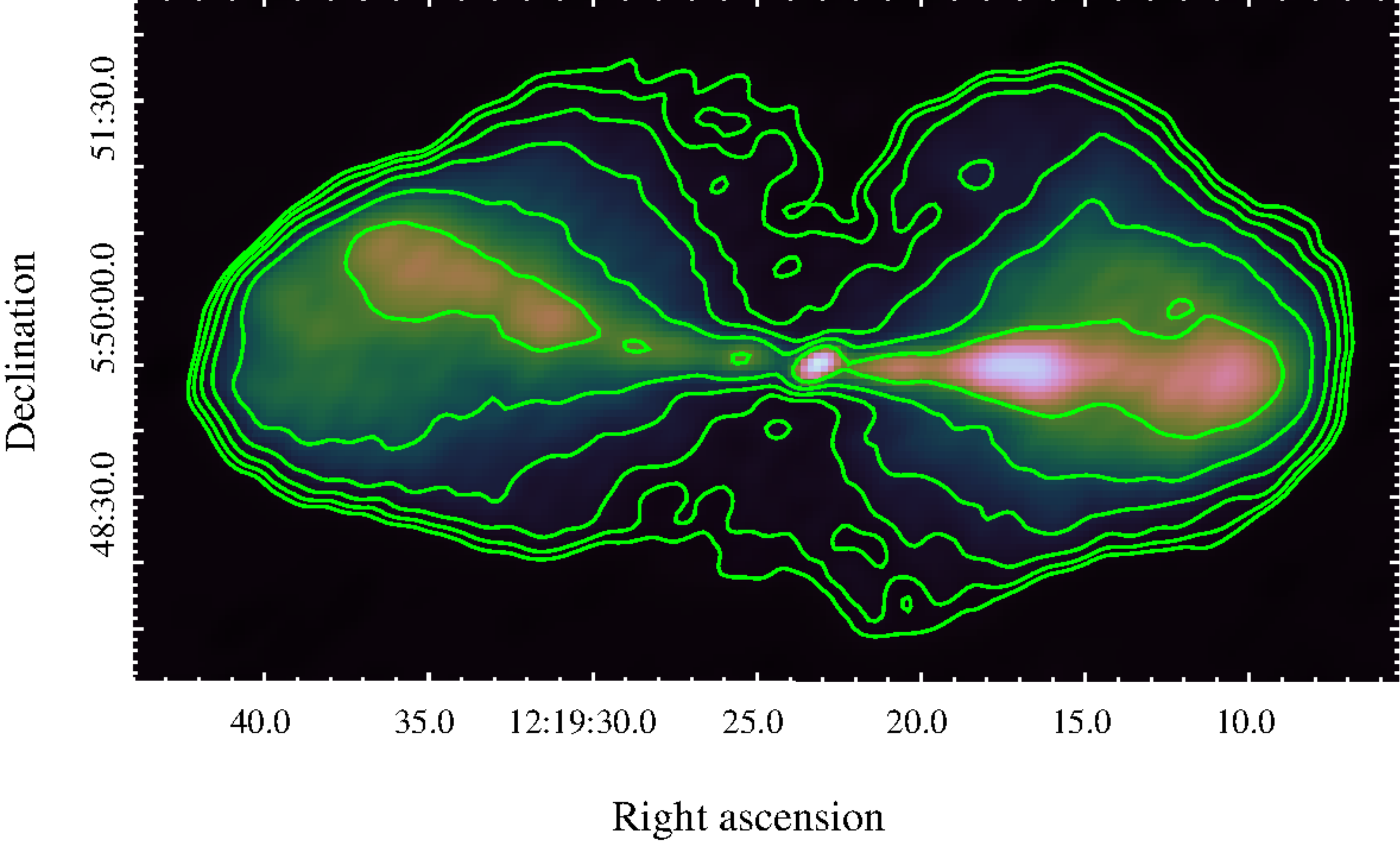}
\caption{\label{fig:1.5GHz}VLA 1.55 GHz radio contours of 3C 270 overlaid on the radio emission. The 1 $\sigma$ rms noise is $\sim$ 0.8~mJy~beam$^{-1}$. Contour levels start at 4$\sigma$ and rise by factors of 2. Flux Density: 18.3 Jy, Restoring beam: $15.78\times10.69$, P.A.= -57.44$^{\circ}$.}
\end{figure}

\begin{figure}
\includegraphics[trim=8.2cm 2cm 11.3cm 0cm, clip=true,width=\columnwidth]{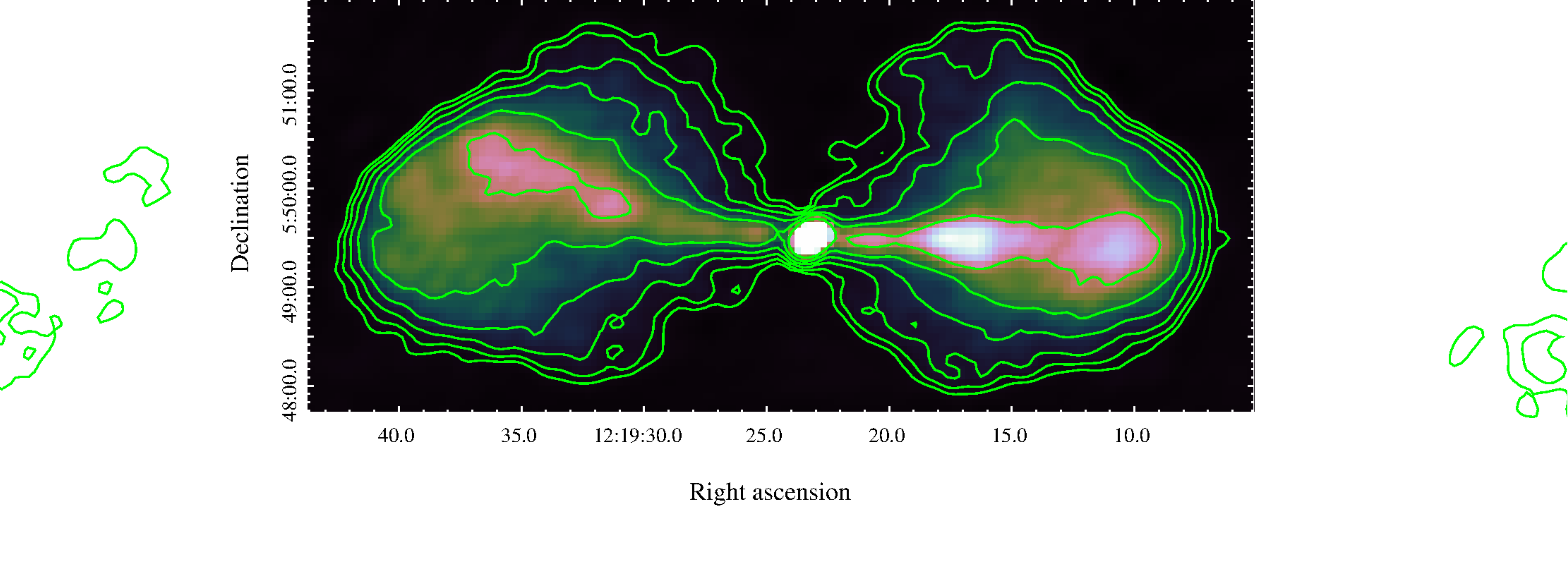}
\caption{\label{fig:5GHz}VLA 4.86 GHz radio contours of 3C 270 overlaid on the radio emission. The 1 $\sigma$ rms noise is $\sim$ 0.4~mJy~beam$^{-1}$. Contour levels start at 4$\sigma$ and rise by factors of 2. Flux Density: 8.3 Jy, Restoring beam: $19.01''\times11.17''$, P.A.= -55.35$^{\circ}$.}
\end{figure}

\section{Radio images}
\label{sec:radims}

Figure~\ref{fig:labels} presents the GMRT 240 MHz radio image of 3C~270 at
a resolution of 15.3$^{''} \times 11.0^{''}$ overlaid on the optical image
from the Digitized Sky Survey. The contour levels start at 4$\sigma$ and
rise by factors of 2. The most prominent components of the source, that is,
the two symmetrical bright jets and their surrounding lobes, are
labelled. The VLA images at 1.55 GHz and 4.86 GHz are shown in
  Figures~\ref{fig:1.5GHz} and \ref{fig:5GHz} respectively. The jets are
  more prominent on the higher-frequency maps.

The radio jets extend in opposite directions for several tens of kpc from the host galaxy of NGC~4261, having a similar morphology at all frequencies and at different restoring beams. The two bright jets are continuous and inflate two fairly round radio lobes that surround the jets. However, as expected, the prominence of the compact core increases with frequency, and it is clearly detected at 1.55 and 4.86~GHz, but not at 240~MHz.

There are two morphological differences between the two jets: first, the eastern jet bends to the north whereas the western one is straight but noticeably broadens and brightens $\sim$1-2$^{\prime}$ from the nucleus; second, the western lobe is slightly less extended than the eastern one. These differences may be caused by interactions with their surroundings, but we see no evidence of a denser environment on either side in the X-ray. \citet{ewan4261} note that the change in the direction of the eastern jet could be misleading, since we are viewing a projected image of a three-dimensional structure.

Table~\ref{tab:radprop} summarises the properties of the source and its components including flux density, the spectral index in the 240 MHz$-$1.55 GHz frequency range and the apparent linear size. The largest projected linear size of the source at all frequencies is $\sim$80 kpc, with a total angular extent of $\sim530''$ along the east/west axis. The jets have similar flux density and size, with the eastern jet being about 2 kpc longer, while the western jet flux density is $\sim$10$\%$ greater than the eastern one. As is evident from Figures~\ref{fig:labels}, \ref{fig:1.5GHz} and \ref{fig:5GHz}, at lower frequencies the lobe emission extends far enough to connect the lobes north and south of the nucleus, but at 4.8~GHz no such connection is observed. This is the first clear indication of the strong spectral gradients present in the radio emission of NGC 4261.

\begin{figure}
\includegraphics[trim=0.7cm 6.5cm 0cm 4.2cm, clip=true, width=\columnwidth]{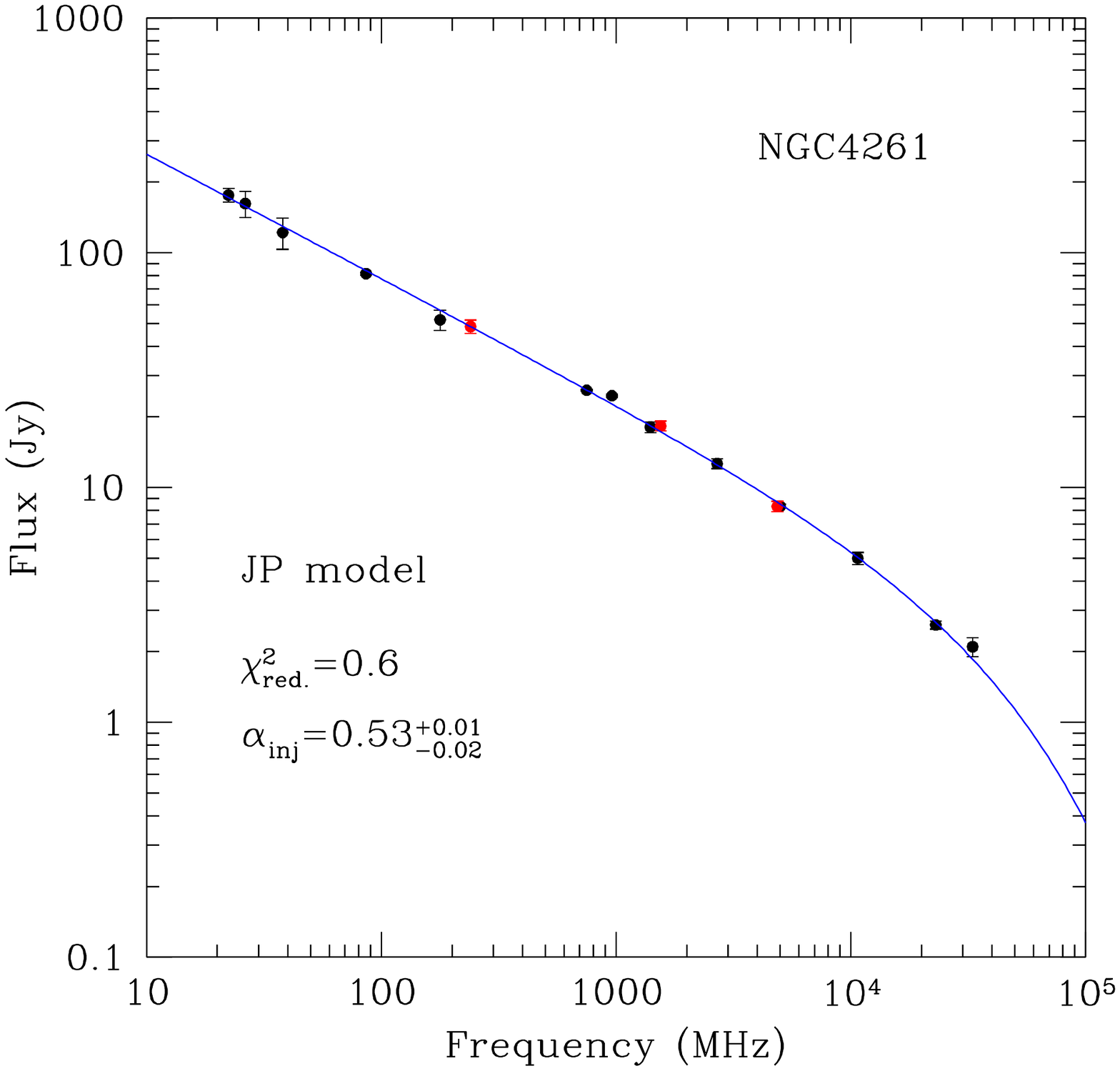}
\caption{\label{fig:spec}Radio spectrum of NGC 4261 between 22 MHz and 33 GHz. The black points are literature data, (see Table~\ref{tab:litflux}), and the red points are the VLA 4.86 GHz and 1.55 GHz, and the GMRT 240 MHz analysed in this paper. The best-fit curve is a Jaffe-Perola (JP) model applied to the data. The best fitting value of the spectral index at injection, $\alpha_{inj}$} is given, as well as the reduced $\chi^{2}$ value from the fit.
\end{figure}

\section{Radio spectral analysis and physical parameters}
\label{sec:analysis}

The archival GMRT and VLA radio data were combined with data from the literature in order to calculate the integrated radio spectrum of 3C~270. The physical parameters of 3C~270 were then calculated using the spectral index distribution of the source. The analysis described below was performed using the Synage++ package \citep{MurgiaPhD}.

\subsection{Spectral analysis}
\label{sec:spec}

The integrated radio spectrum between 22 MHz and 33 GHz for 3C~270 is shown
in Figure~\ref{fig:spec}. Table~\ref{tab:litflux} lists the
  literature flux densities used in the spectrum, and the source of each
  measurement. Adding our VLA measurements to the literature data, we fit
  a \citep{JaffePerola73} (JP) model to the integrated spectrum. Our aim in
  fitting this model is firstly to determine whether the low-frequency data
  can be well modelled as a power-law, whose slope will approximate the
  injection spectral index, and secondly to allow us to compare our GMRT
  flux density measurements with the well-calibrated data from other
  frequencies. The model assumes that the radiative timescale of the
  electrons is much longer than their timescale for continuous
  isotropisation. The resulting fit describes the low frequency data well,
  with an asymptotic low-frequency index of
  $\alpha_{inj}=0.53_{-0.02}^{+0.01}$, typical of FR~I jets
  \citep{Young05}. The slope steepens above $\sim$5~GHz, with a
  break frequency of $\sim$85 GHz, though this is likely driven to a high
  value owing to the high luminosity of the relatively young plasma in the
  jets and core. We note that a continuous-injection (CI) model gives an
  identical fit.

\begin{figure*}
\includegraphics[width=\textwidth]{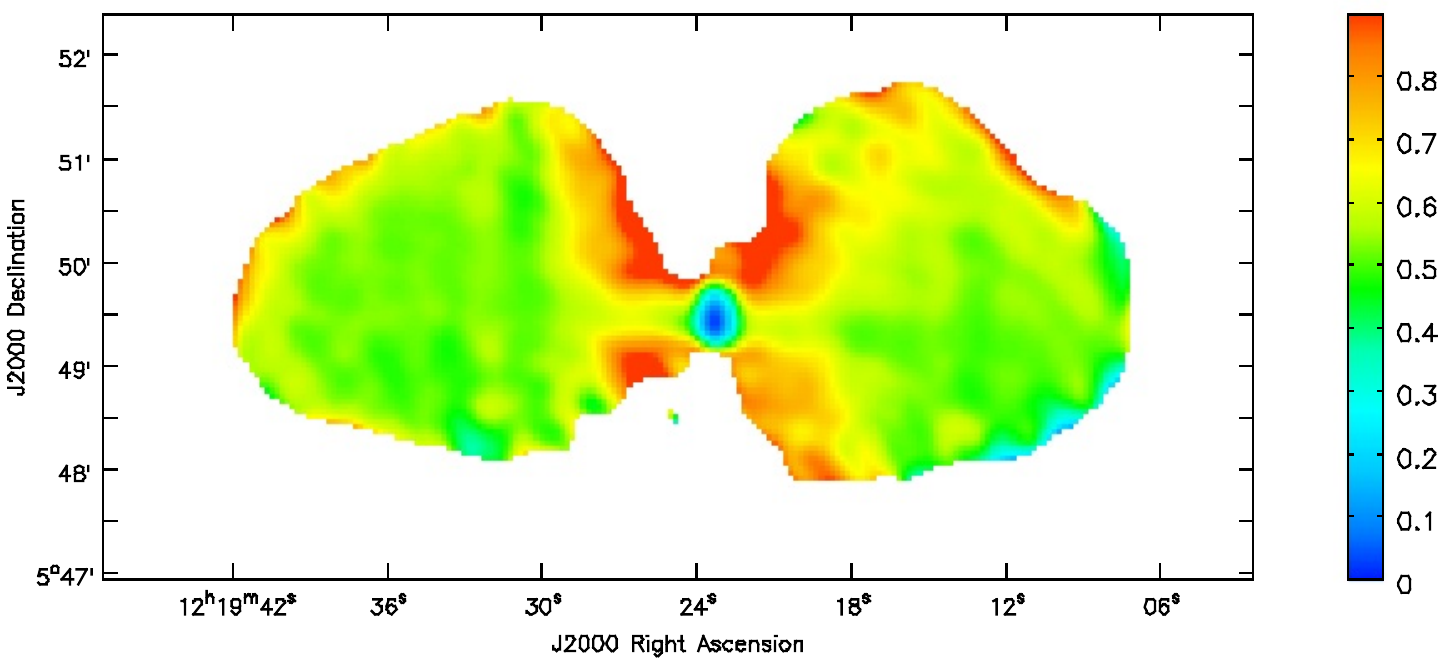}
\caption{\label{fig:spixG}Map of the spectral index distribution between 240 MHz and 1.55 GHz based on images with 20$^{''}$ HPBW. The typical uncertainty on spectral index in the lobes is $\pm$0.05. The values within 1 beam of the sharp edges of the lobes are likely to be unreliable.}
\end{figure*}

\begin{figure*}
\includegraphics[width=\textwidth]{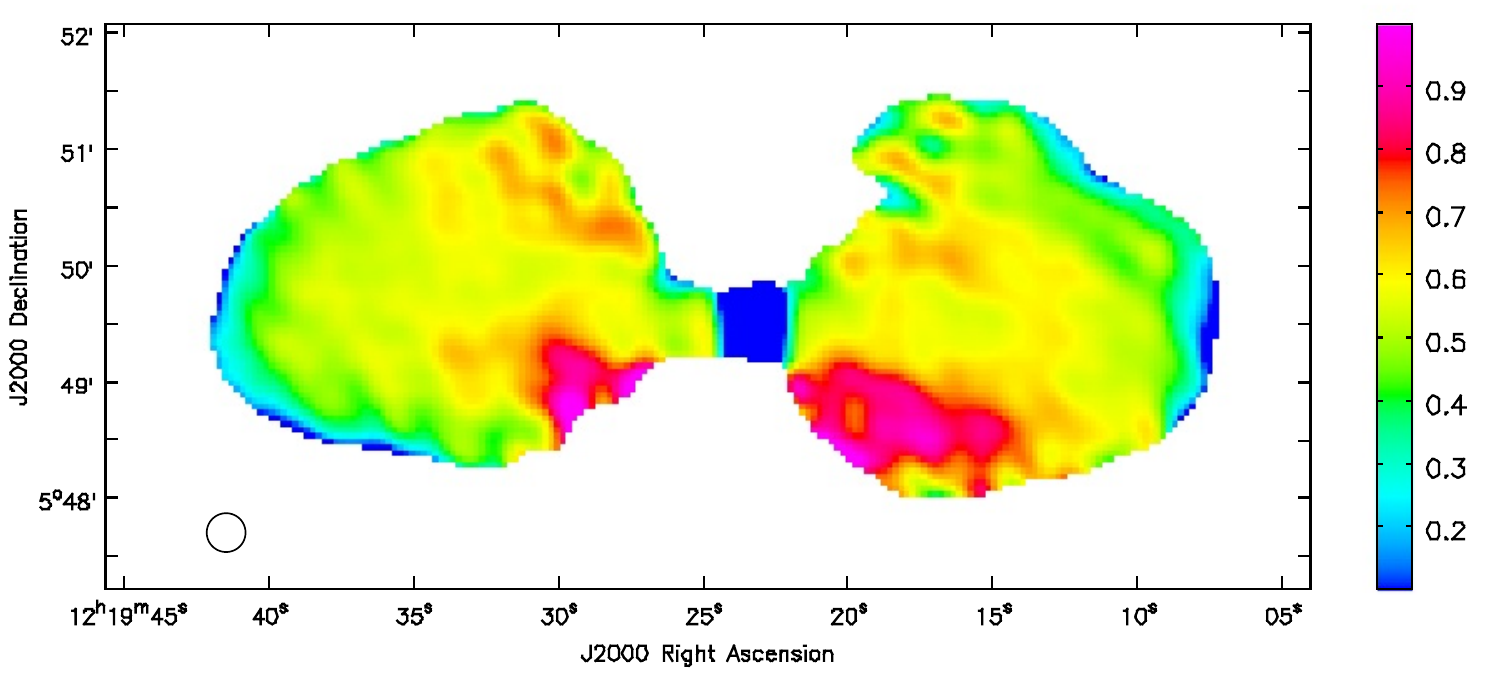}
\caption{\label{fig:spixV}Map of the spectral index distribution between 1.55 GHz and 4.8 GHz based on images with 20$^{''}$ HPBW (indicated by the circle on the lower left). The typical uncertainty on spectral index in the lobes is $\pm$0.06. The values within 1 beam of the sharp edges of the lobes are likely to be unreliable.}
\end{figure*}

The maximum spatial scale resolved in our GMRT and VLA images is large compared to the source, with the possible exception of the 4.86~GHz VLA observation ($\sim$72$^{\prime}$, 28$^{\prime}$, 19$^{\prime}$ and 6$^{\prime}$ at 240, 610, 1550 and 4860~MHz respectively). However, the VLA 1.55 and 4.86 GHz data points lie close to the fit and are both in good agreement with previous measurements at these frequencies, indicating that the VLA data capture the great majority of the flux from the source at both frequencies.

The GMRT data points are in good agreement with the model fit, suggesting that the 20\% correction we applied to the GMRT flux calibration is correct. Without this correction, the GMRT 240~MHz flux density would fall $\sim$2.8$\sigma$ below the model prediction. Including systematic uncertainties on the flux scale and the elevation-dependent error on GMRT flux calibration would bring the measurement to within 1$\sigma$ of the model. However, we note that all the bright sources in the GMRT fields for which multi-wavelength data are available show systematically low flux densities; while the difference may not be strongly significant at a single frequency for 3C~270, the agreement across multiple sources and both bands suggests that the problem is real and significant.

To provide a more nuanced picture of the spectral index structure
  within 3C~270, we created spectral index maps of the source in two
  frequency ranges, 240~MHz-1.55~GHz and 1.55-4.86~GHz. Each set of images
  was produced in IMAGR using the same \emph{uv} range, Gaussian taper and
  cellsize. We used a 20$^{''}$ circular restoring beam (chosen based on
  the lowest resolution 1.55~GHz image), corrected for the primary beam,
and aligned the images. Regions where the flux density in either
  band was $<$5$\sigma$ significant were removed. The resulting spectral
  index maps are shown in Figures~\ref{fig:spixG} and \ref{fig:spixV}. Typical uncertainties on the spectral index in the lobes (excluding systematic uncertainties on the flux density scale) are $\pm$0.06 for the 1.55-4.8~GHz map and $\pm$0.05 for 240~MHz-1.55~GHz. Uncertainties in the spectral indices of the jets and core can be found in Table~\ref{tab:radprop}.

In Figure~\ref{fig:spixG}, the unresolved core of the radio source can be seen. It has a spectral index $\alpha_{240}^{1550}\sim0.16\pm0.10$, probably as a result of self-absorption (as indicated by the lack of an obvious core in the 240 MHz image, Fig.~\ref{fig:labels}) since the spectral index is rather frequency-dependent (high curvature). Moving outwards the spectral index becomes steeper and a constant value of $\alpha\sim0.47\pm0.09$ can be seen in the jets, while the lobes that dominate the total index in the lower-frequency band appear with constant but slightly steeper values ($\alpha\sim0.60\pm0.09$). In Figure~\ref{fig:spixV} the core is obvious with a flat spectral index $\alpha_{1550}^{4860}\sim0.03\pm0.08$. The steepening of the spectral index within the lobes is much clearer in this image as a spectral index of $\sim0.9 - 1$ can be seen above and below the beginning of the jets. In the jets, the spectral index at the point where they are surrounded by the lobes is $\sim0.5 - 0.6$ and remains constant throughout. The spectrum in the lobes gradually steepens towards the nucleus, reaching values of $\sim$1$\pm$0.06. In both Figures 5 and 6, the features in the spectral-index maps at the edges of the source, and particularly at the ends of the lobes, are likely to be spurious and induced by residual reconstruction errors in regions of steep intensity gradient.

\begin{figure}
\includegraphics[width=\columnwidth]{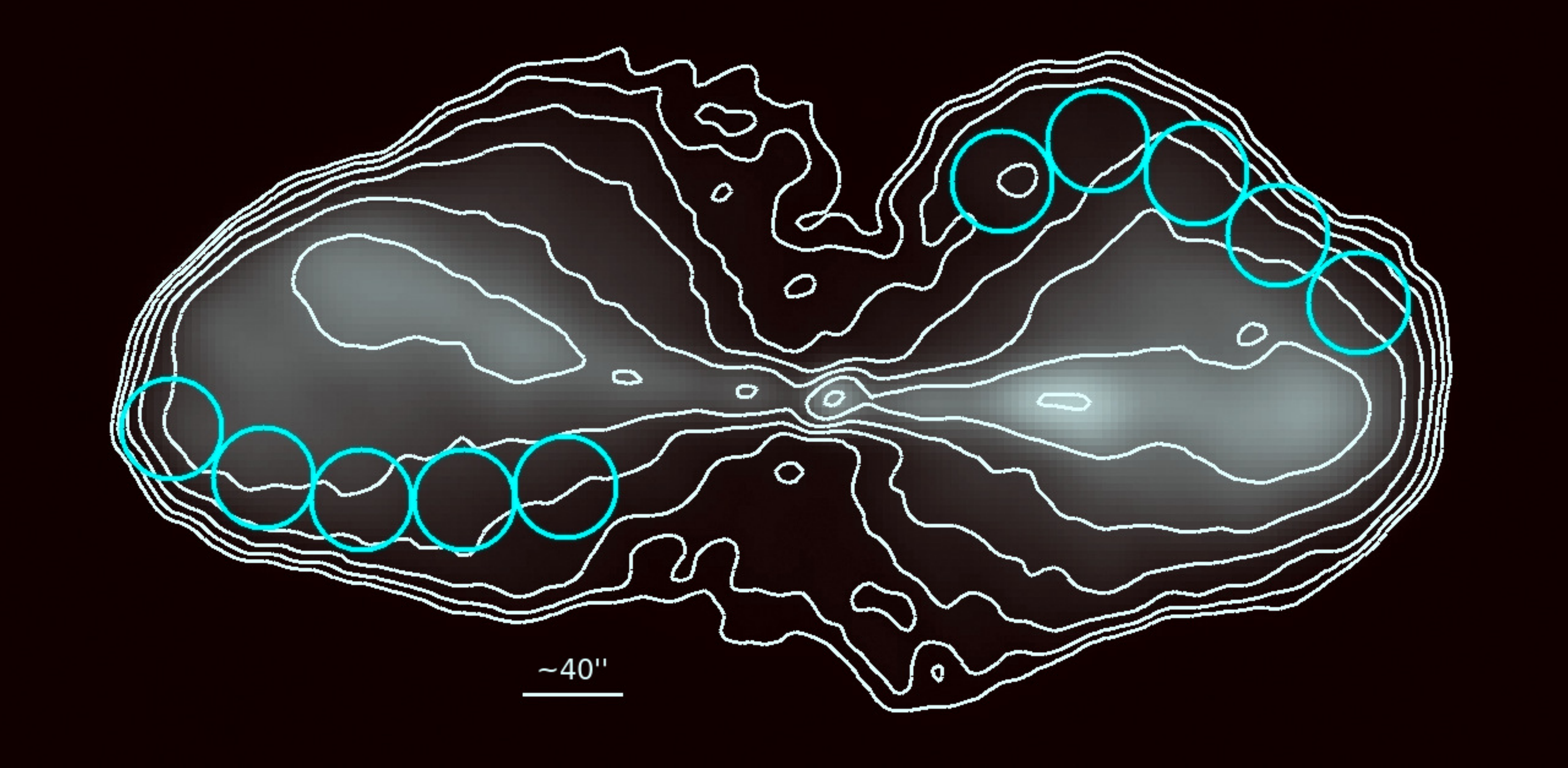}
\caption{\label{fig:profreg}1.55 GHz image in grey scale. Overlaid are the 1.55 GHz contour levels starting at 3.2~mJy~beam$^{-1}$ and rising by factor of 2. The cyan circles indicate the lobe regions used for the source age estimation using the spectral index profile method.}
\end{figure}

\begin{figure}
\includegraphics[trim=1.5cm 6.5cm 0cm 3.8cm, clip=true, width=\columnwidth]{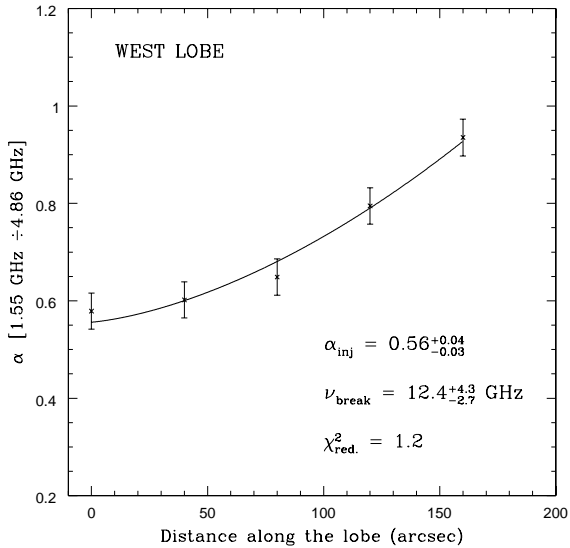}
\caption{\label{fig:spixprofW}Spectral-index distribution along the west lobe of the source calculated from the 1.55 and 4.86 GHz frequencies, derived using the circular regions shown in Fig.~\ref{fig:profreg}. The solid line represents the best JP model fit. The values of $\alpha_{inj}$ and $\nu_{break}$ along with the reduced $\chi^{2}$ from the fit are reported here. }
\end{figure}

\begin{figure}
\includegraphics[trim=1.5cm 6.5cm 0cm 3.8cm, clip=true, width=\columnwidth]{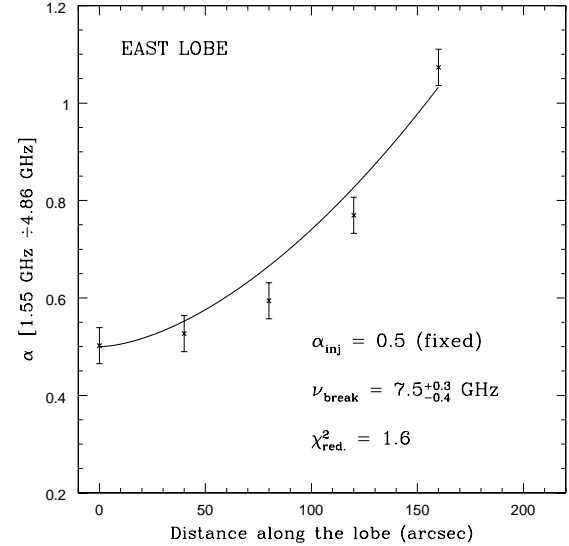}
\caption{\label{fig:spixprofE}As for Figure~\ref{fig:spixprofW} but for the east lobe.}
\end{figure}

We performed a fit of the observed spectral index trend through the lobes using the techniques of \cite{Murgia03}. As Figures~\ref{fig:spixG} and \ref{fig:spixV} show, $\alpha_{1550}^{4860}$ changes much more through the lobes than $\alpha_{240}^{1550}$, indicating that the break frequency is $\gtrsim$1.55~GHz over most of the area. The Synage++ package only considers spectral indices between two frequencies, and we therefore use it to model the trend in $\alpha_{1550}^{4860}$, while confirming that the 240~MHz data at each position do not conflict with the higher frequency data. The spectral index trend was determined by averaging the flux densities of each frequency in 40$^{''}$ diameter circular regions starting in the region likely to contain the youngest plasma, at the end of the jets, and then moving back along the paths shown in Figure~\ref{fig:profreg} to the oldest plasma toward the nucleus. The circular regions were chosen to be larger than the beam size so that the measurements of the spectral index are independent. Figures~\ref{fig:spixprofW} and \ref{fig:spixprofE} show the resulting spectral index measurements in the west and east lobes respectively.

The trends in the west and east lobe were fitted using a JP model. Assuming
a simple constant source expansion velocity ($d\propto t$), and
  knowing that the break frequency of the emission from a passively ageing
  relativistic plasma $\nu_{break}\propto$~1/age$^2$, we expect to find
  $\nu_{break} \propto d^{-2}$, where $d$ is the distance from the end of
  the jet. The best fit for the observed spectral trend in west and east
lobe of the source can be seen as a solid line in
Figures~\ref{fig:spixprofW} and \ref{fig:spixprofE}, respectively.
We note that the model provides a reasonable description of the
  spectral index trend in both lobes, suggesting that our assumption of
  constant expansion velocity is not unreasonable. The model provided
similar values of $\alpha_{inj}$ and $\nu_{break}$ for both lobe regions.
In the west lobe the calculated injection spectral index from the model
was~$\alpha_{inj}\sim0.56$ with a break frequency of $\nu_{break}\sim12.4$
GHz while in the east lobe we used a fixed injection spectral index of
$\alpha_{inj}=0.5$ and got a $\nu_{break}\sim 7.5$ GHz (see also
Table~\ref{tab:SGfit}). The break frequencies measured change
  little if the injection indices in the two lobe heads are assumed to be
  the same, either 0.5 or 0.56, and we note that both values are similar to
  the value of 0.53 found from the integrated spectrum. The similarity
  between the two lobes indicates that they have similar plasma properties,
  and we therefore consider it reasonable to fix $\alpha_{inj}$ in the east
  lobe.

\begin{table}
\begin{center}
\caption{\label{tab:litflux}Integrated radio fluxes for NGC~4261 drawn from the literature}
\begin{tabular}{lcc}
\hline
 $\nu$         &  Flux Density & Reference \\
    (MHz)        &   (Jy)        & \\
\hline
 22.3            & 176 $\pm$ 12     & 1       \\ 
 26.3            & 162 $\pm$ 21     & 2       \\ 
 38              & 144 $\pm$ 7.2      & 3        \\ 
 86              & 81.5 $\pm$ 1.1     & 4        \\ 
 178             & 56.5 $\pm$ 2.8       & 3        \\ 
 750             & 27.15 $\pm$ 0.23       & 5         \\ 
 1400            & 18.62 $\pm$ 0.93       & 3           \\ 
 2695            & 12.79 $\pm$ 0.64       & 3            \\ 
 5000            & 8.26 $\pm$ 0.15        & 6           \\ 
 10700           & 4.69 $\pm$ 0.28        & 7          \\ 
 23000           & 2.6 $\pm$ 0.1        & 8             \\ 
 33000           & 2.1 $\pm$ 0.2        & 8              \\
\hline
\end{tabular}
\end{center}
REFERENCES.-(1) \cite{Roger69};  (2) \cite{Viner75};  (3) \cite{Kellermann69};  (4) \cite{Artyukh69};  (5) \cite{PaulinyToth66}; (6) \cite{PaulinyToth68}; (7) \cite{Kellermann73};   (8) WMAP, \cite{Bennett03}
\end{table}


\begin{table}
\begin{center}
\caption{\label{tab:SGfit}Results of the spectral index profile analysis. Values marked $^*$ were fixed during fitting.}
\begin{tabular}{lcccc}
\hline
 Region  &  $\alpha_{\rm inj}$ &  $\nu_{\rm break}$  & $\tau_{\rm rad}$ &    $\upsilon_{\rm adv}$  \\
         &                 &   (GHz)         & ($10^{6}$ yrs)&                      \\
\hline
 West lobe       & $0.56_{-0.03}^{+0.04}$ &  $12.4_{-2.7}^{+4.3}$  &   29$_{-4}^{+3}$   &     $\sim0.004c$      \\ 
 East lobe       & $0.50^*$    &   $7.5_{-0.4}^{+0.3}$  & 37$_{-1}^{+2}$    &   $\sim0.004c$          \\ 
\hline
\end{tabular}
\end{center}
\end{table}

\subsection{Physical parameters of 3C 270}
\label{sec:phys}

To calculate the key physical parameters of 3C~270, we have to make certain assumptions. Firstly, the relativistic particle and magnetic field energy densities were assumed to be uniformly distributed over the volume occupied by the radio source and in approximate energy equipartition. A low energy cut-off $\gamma_{min}$ (where $\gamma$ is the electron Lorentz factor) was introduced in the energy distribution of the radiating electrons rather than the fixed 10 MHz $-$ 100 GHz frequency interval used in the classic equipartition equations \citep{Pacholczyk70}. We assumed that the magnetic field is unordered along the line of sight and used $\gamma_{min}=100$ in the electron energy distribution, which corresponds to $\sim50$ MeV. 

\begin{figure}
\includegraphics[width=\columnwidth]{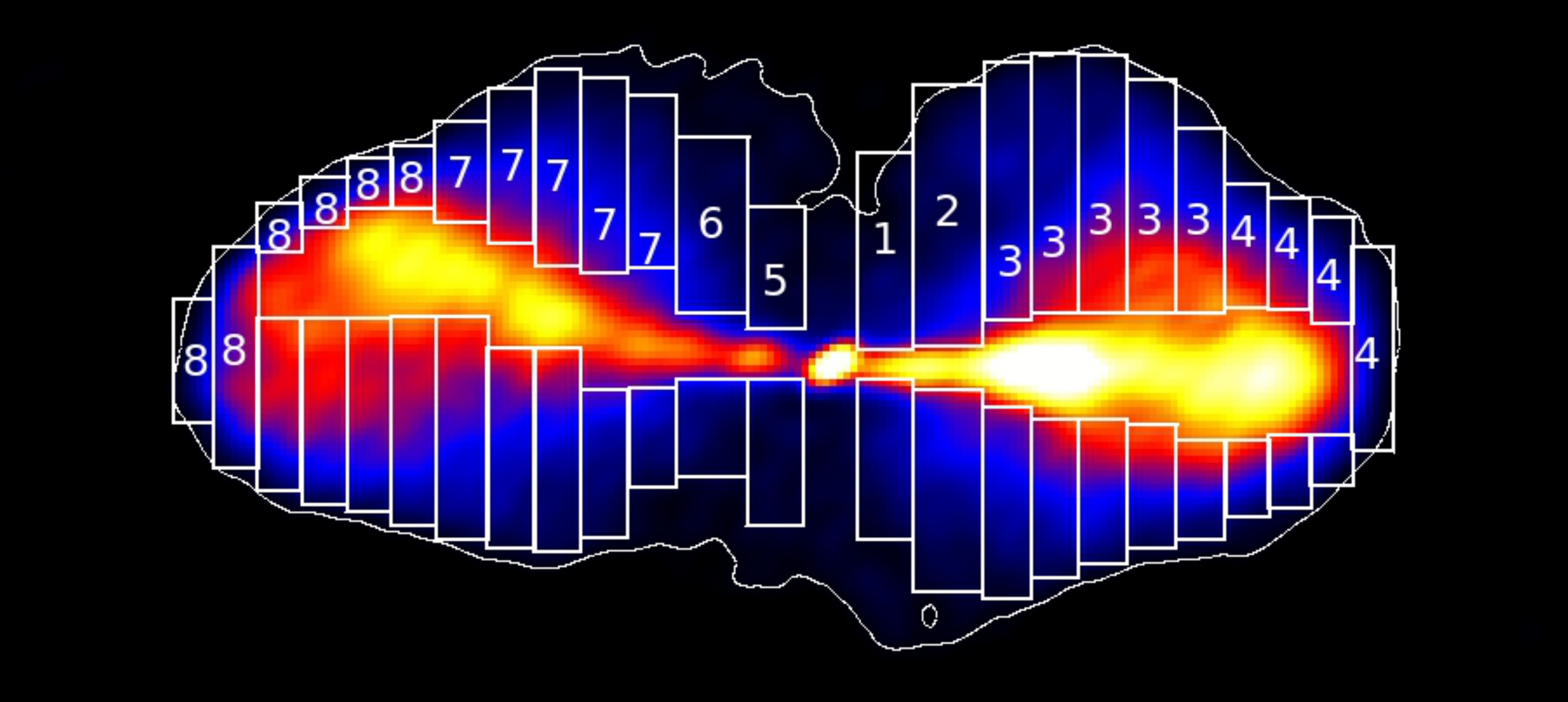}
\caption{\label{fig:vol}Regions used for the estimation of the volume and the other physical parameters described in Table~\ref{tab:phys}. The 3.2~$\mu$Jy (4$\sigma$) contour is shown to indicate the limit of the detected lobe emission.}
\end{figure}

We used the injection spectral index derived from the west lobe $\alpha_{inj}$=0.5, since this is in broad agreement with the value for the east lobe and the source as a whole. The flux density at 240 MHz (Table~\ref{tab:phys}) was used to calculate the radio luminosity of the source, since among the frequencies used in this work it is least affected by electron spectral ageing. As the lobes are not well described by simple ellipsoids, we divided them into a number of rectangular regions chosen to approximate the breadth of the lobe at a range of distances along the jet axis. These are shown in Figure~\ref{fig:vol}. Excluding the jets from the equipartition calculations for the lobes, we assume rotational symmetry about the jet axis to estimate the volume in each cylindrical region. We then group the regions, numbering these groups 1-8, to match the radial binning of the IGM pressure profile derived from X-ray observations. We also assume that the axis of the radio galaxy is in the plane of the sky. If the jet axis is oriented at 76$^{\circ}$ to the plane of the sky \citep[as suggested by the most recent modelling,][]{Laing14}, our estimated volume will increase by $\sim$10\%, altering our estimate of equipartition magnetic field by $\sim$10\%.

The volumes, flux densities, spectral indices and equipartition magnetic field in regions 1-8 are listed in Table~\ref{tab:phys}. We also estimated the pressure of the relativistic plasma in each region, which is defined to be

\begin{equation}
P_{radio}=\frac{B^2_{eq}}{2\mu_0\epsilon}+\frac{(1+k)E_e}{3V\phi},
\label{egn:Prad}
\end{equation}

where $E_e$ is the energy of the electron population, $V$ is
  the volume, $\mu_0$ is the permeability of free space (4$\pi$ in cgs
  units or 4$\pi \times 10^{-7}$ in SI), $\phi$ is the filling factor of
  the plasma (assumed to be 1) and $k$ is the energy ratio of the
  non-radiating particles to the radiating relativistic electrons. We
  assume $k$=1, indicating equal energy in radiating and non-radiating
  particles. The factor $\epsilon$ describes the ordering of the magnetic
  field and we adopt $\epsilon=3$, indicating a tangled magnetic field, as
  in \cite{ewan4261}.

We find $B_{eq}\sim$5~$\mu$G for both eastern and western lobes, though the outer parts of the lobes (regions 4 and 8) have higher $B_{eq}$, close to the estimate for the source as a whole $\sim$6.4~$\mu$G. For this field (see later), electrons at $\gamma_{min}$=100 radiate at $\sim$5 MHz. 

\begin{table*}
\begin{center}
\caption{\label{tab:phys}Physical parameters of 3C~270 and its components}
\begin{tabular}{lccccccccc}
\hline
 Region   &  r$_{\rm min}$ &  r$_{\rm max}$  &  V$_{\rm tot}$ &    $S_{240 \rm MHz}$   &  $S_{1550 \rm MHz}$  & $\alpha_{\rm obs}$ &  $L_{\rm 1550}$ & $B_{\rm eq}$ & $P_{\rm radio}$ \\
          & (arcsec)   &  (arcsec)  &  (kpc$^{3}$)&      (mJy)         &    (mJy)         &                &   (10$^{22}$ W/Hz)   & ($\mu$G)  & ($10^{-12}$dyn cm$^{-2}$) \\
\hline
West Lobe &              &          &             &                   &                    &                  &            &          &               \\
  1       &      10     & 33       &  1724         &  $998\pm80$       &      $316\pm16$    &    0.62          &   3.6       &   5.0     &  1.42     \\
  2       &    33       &  63      &   3861         &    $2067\pm165$     &  $650\pm33$      &      0.62       &     7.3    &    4.9     &  1.36   \\ 
  3       &     63     &     168   &     12081       &   $8741\pm700$     &    $3138\pm157$   &       0.55     &    35     &     5.5     &1.73      \\ 
  4       & 168        &     242   &    2551       &        $2262\pm181$   &  $897\pm45$      &      0.50     &      10    &     6.0   &   2.06 \\ 
   
\hline
East Lobe &           &               &             &                     &                    &           &          &           &      \\
  5      &  10        &      33       &      1056     &    $408\pm33$       &    $146\pm7$       &     0.55  &  1.6     &   4.6     &  1.21   \\ 
  6      &   33       &    63         &    1646       &      $976\pm78$      &    $287\pm14$     &    0.66   &  3.2     &    4.9    &  1.39     \\
  7      &     63     &      168      &      9753      &     $5868\pm469$    &      $2310\pm116$ &  0.50     &  26      &    5.3   &  1.64   \\ 
  8      &    168     &      282      &     4668       &    $5755\pm460$     &      $2202\pm110$ &  0.52     &  25      &    6.5   &  2.43   \\ 
\hline
\end{tabular}
\end{center}
\end{table*}


\subsection{Radiative age}

When considering the development of the radio jets and lobes, we adopt a simple conceptual model, in which at any given time plasma is rapidly transported out from the central engine to the tip of the jet. At this point the jet flow is halted by contact with the IGM, and the plasma driven out of the jet moves out radially to form the lobe. In this model, the age of the lobe plasma is linked to the distance along the jet; the back of the lobes (close to the nucleus) formed when the jet only extended a few kiloparsecs, while the tips of the lobes have only formed recently. In principle, backflows could develop, transporting younger plasma away from the jet tips into regions of the lobe formed at earlier times. However, in the absence of such a flow, or external forces causing reacceleration, plasma in the lobes is expected to age passively, with the high-frequency break in its spectral index falling to lower frequencies over time.

The radiative age of the radio source of NGC 4261 can be calculated if the break frequency of the source is known \citep[e.g.,][]{Myers84}. The magnetic field is considered to be uniform and constant over the whole of the source lifetime when radiative losses dominate over expansion losses. Neglecting reacceleration processes, the radiative age (the time since the last period of injection of relativistic electrons began) can be obtained as 

\begin{equation}
 t_{\rm rad}=1590\frac{B^{0.5}_{\rm eq}}{B^{2}_{\rm eq}+B^{2}_{\rm CMB}}[(1+\emph{z})\nu_{\rm break}]^{-0.5}\ {\rm Myr},
\label{eqn:t_rad}
\end{equation}
where $\nu_{\rm break}$ is expressed in GHz, and $B_{\rm eq}$ and $B_{\rm CMB}$ in $\mu$G \citep{Parma07}. $B_{\rm eq}$ is the equipartition magnetic field and $B_{\rm CMB}=3.2$ is the equivalent magnetic field strength of the cosmic microwave background (CMB) radiation at zero redshift, and equation~\ref{eqn:t_rad} thus includes both inverse Compton and synchrotron losses.

Using equation \ref{eqn:t_rad} and the mean magnetic field for the source, 5~$\mu$G, we estimated a radiative age of $\sim$29 Myr for the western lobe and $\sim$37 Myr for the eastern one. We also derived an estimate of the advance speed of the lobe tip, $\upsilon_{\rm adv}$/c, where c is the speed of light. This velocity was calculated as $\upsilon_{\rm adv}$= LD/$\tau_{\rm rad}$, assuming a simple constant velocity and using the linear distance (LD) from the back of the lobe to the tip of the jet as measured from the radio images in Section~\ref{sec:radims}. We found that for both lobes $\upsilon_{\rm adv}\sim$ 0.004c ($\sim$1300 km s$^{-1}$ for the western lobe and $\sim$1200 km s$^{-1}$ for the eastern one).

By comparison, the sound speed in the IGM in the environment around NGC~4261, estimated from the measured X-ray temperature, increases from about 380 km s$^{-1}$ near the core to 590 km s$^{-1}$ beyond $\sim$100$''$. The advance speed of the lobe tip therefore appears to be supersonic relative to the IGM. Conversely, the Alfv\'{e}n speed in the lobe is about 10$^5$ km~s$^{-1}$ (using the lobe plasma properties deduced in Section~\ref{sec:press}). That is, speed of expansion of the lobe is strongly sub-Alfv\'{e}nic, and so will not excite magnetic shocks in the lobes. The internal characteristic speed of the waves in the magnetized relativistic plasma were calculated from


\begin{equation}
 \upsilon_{A}=\frac{c}{\sqrt{1+\frac{u_{ptcl}+P_{radio}}{2P_{mag}}}}
\label{eqn:alfven}
\end{equation}
where $\upsilon_A$ is the Alfv\'{e}n speed expressed in m s$^{-1}$, $u_{ptcl}$ is the energy density of the relativistic particle population, and $P_{mag}$ is the magnetic field pressure. 

\section{Discussion}
\label{sec:discuss}

\subsection{The age of the source}
\label{sec:age}

Before considering the implications of our radiative age
  measurements, we first discuss potential sources of error in our
  approach. We have performed a relatively simple modelling of the
  1.55-4.86~GHz spectral index trend in the lobes, and our model rests on a
  number of assumptions. These include constant source
  expansion velocity, constant magnetic field, our choice of spectral
  model, and that reacceleration, adiabatic expansion and mixing are not
  important processes in the lobes. As mentioned in Section~\ref{sec:spec},
  the fact that our model provides a reasonable description of the spectral
  index trend suggests that the assumption of a constant expansion velocity
  is at least reasonable; we see no evidence that velocity changes are
  required, though of course other factors could disguise deviations from
  the trend. As a further test, we refitted the spectral index profiles,
  adding one additional region closer to the AGN. In both lobes, the fits
  were poorer, primarily because of the differences in the extent of the
  emission seen in the lobes close to the nucleus (back of the lobes) at 1.55 and 4.8~GHz, which lead
  to divergence from the smooth trend in spectral index. For example, in
  the west lobe, the added region is close to the edge of the lobe at
  4.8~GHz, leading to a steeper than expected spectral index and a reduced
  $\chi^2$=8.7. Both fits find lower break frequencies $\sim$4-5~GHz,
  suggesting radiative ages $\sim$45~Myr, older than our best estimate, but
  not by a large factor.

Our use of the JP model may be over-simple, but we are limited in this by the available data. \citet{Harwood13} compare the utility of several spectral models to describe spectral ageing in FR-II galaxies and conclude that the \citet{Tribble93} model provides the best combination of fit quality and physical realism; Harwood et al. consider the JP model at least as physically accurate, but it provides a poorer fit. However, they find that to differentiate between models requires data at 7-9 frequencies, whereas we have only 3 available. Further observations are therefore needed to determine whether a different model might provide a better description (and age estimate) for 3C~270.

Losses caused by adiabatic expansion would affect our age estimate, lowering the break frequency and therefore the estimated age. However, adiabatic losses decrease the luminosity of the plasma, and are more effective in doing so than they are in reducing $\nu_{break}$. We might expect losses in the jets as they flare from the narrowly collimated structures seen in the X-ray to the broad jets we observe in the radio, but the fact that the lobes have high surface brightnesses and smooth surface brightness distributions suggests that adiabatic losses within the lobes are probably not important. Variation in the magnetic field within the lobes could affect our results, but on large scales we only measure a relatively small ($\sim$20\%) change in field strength across the lobes (see Table~\ref{tab:phys}), perhaps as little as 8\% in the radius range used to estimate age. This is insufficient to balance the observed spectral index trend, and we therefore consider our assumption of a constant field to be a reasonable approximation. Small-scale field variations cannot be ruled out, but would again require improved spectral modelling and additional observations.

Moving on to the results, our spectral-index profile modelling
(Figures~\ref{fig:spixprofW} and \ref{fig:spixprofE}), implies ages of $t_{\rm rad}\sim29$ and 37~Myr for the west and east lobes respectively (Table~\ref{tab:SGfit}). These ages are less than the $<$75~Myr upper limit estimated by \citet{ewan4261} from the IGM sound speed. While the upper limit and our values are consistent, the factor two difference between the values seems to be significant. \citet{Eilek96} first noted that the synchrotron ages for FRI and FRII radio sources derived from the spectra differed from the dynamical ages as much as a factor of 10. By this standard the factor of 2 discrepancy we observe is relatively mild. Several studies suggest plausible explanations for the age difference between spectral and dynamical ages \citep{Blundell00}, including in situ acceleration of particles \citep[e.g.,][]{Eilek96,Carilli91} and magnetic fields that vary with position in the lobes \citep[e.g.,][]{Katz93,EilekArendt96}. Studies have shown much better agreement between dynamical and spectral ages in young sources ($<$10~Myr, \citealt{MatthewsScheuer90,Kaiser97}), but the cause of the discrepancy in older sources is still unknown. We will discuss three possible explanations for the age discrepancy in 3C~270: i) a period of highly supersonic expansion early in the AGN outburst; ii) multiple outbursts or large changes in jet power; iii) a backflow capable of mixing younger electrons from the jets into the old plasma at the back of the lobes.

In scenario (i), the expansion of the lobes was much faster in the past, so that the assumption of a constant near-sonic expansion velocity by \citet{ewan4261} leads to an overestimate of the age of the source. For the larger eastern lobe, the radiative age (37~Myr) implies that the lobes should be driving into the IGM at Mach$\sim$2, if they have expanded at constant velocity. Shocks in the IGM, driven by this expansion, would be detectable in the available X-ray observations as increases in temperature and surface brightness, and are ruled out by the measurements described in \citet{ewan4261}. However, there is also the possibility of that a period of supersonic expansion occurred early in the life of the source, followed by a period of slower expansion which we are now observing. In this case, the shocks would have detached from the lobe tips as the expansion speed dropped, and moved outwards, potentially passing out of the field of view of the \textit{XMM-Newton} observation. Assuming that the shock front is just outside the \textit{XMM} field of view ($\sim$14$^\prime$ or $\sim$130~kpc) we can calculate the duration of the supersonic phase. The minimum Mach number of such a shock would be $\cal M_{\rm min}\sim$6 ($\sim$3500~km~s$^{-1}$). The corresponding maximum period of supersonic expansion would be $\sim$7.5~Myr, during which time the lobes would expand to $\sim$28.5~kpc along the jet axis. After this, expansion would continue at the sound speed for the remaining 3/4 of the lifetime of the radio source.

A Mach$\sim$6 shock in the IGM would cause an increase in X-ray
  surface brightness of a factor $\sim$13.5 and a temperature increase of a
  factor $\sim$12. Even if the efficiency of energy deposition is low, such
  a shock would have strongly heated the IGM, and the observed temperature
  would be the result of this heating. For a 10\% heating efficiency, we
  would expect the shock to have deposited $\sim$10$^{58}$~erg in the
  central 10~kpc alone, and the current rate of radiative energy loss,
  $\sim$10$^{41}$~erg~s$^{-1}$ would only have removed $\sim$1\% of this
  energy. This implies a pre-shock core temperature as low as 0.1~keV,
  lower than is observed in any group or cluster, and more comparable with
  the temperature of the interstellar gas in our own Galaxy than that
  expected in a giant elliptical. Since NGC~4261 appears to be fairly
  typical of the population of group and cluster-central FR-I radio
  galaxies, we might also expect to see a population of such sources
  driving shocks of similar strength, with evidence of strong shock
  heating.  No such population has been identified. The scenario of a
  single, highly supersonic outburst therefore seems unlikely.

In scenario (ii) the AGN would have undergone more than one outburst, or the jet power may have varied significantly over time. In this case our estimated age could correspond only to the newest outburst or period of high jet power. A restarted jet, or period of more powerful outflow, could drive shocks into the lobes, causing in situ reacceleration in the old plasma while producing new expansion at the lobe tips. \citet{Worrall10} note structure that could be caused by variations on $\sim$10$^4$~yr timescales in the X-ray jet, while the changes in radio brightness along the western jet suggest variation on longer timescales. This scenario is plausible, and difficult to rule out with current data.

In scenario (iii) a backflow would transport young, energetic
electrons back towards the nucleus, reducing the mean age of the plasma.
  Numerical simulations \citep[e.g.,][]{PeruchoMarti07}
  support the possibility of mildly relativistic backflows in a sheath
  surrounding light FRI jets. Modelling of the base of the jets in 3C~270
  does not reveal any evidence of such a backflow \citep{Laing14}, but it
  cannot be ruled out on larger scales. However, if backflows are limited
  to a sheath around the jets \citep{LaingBridle12}, the question of how electrons are mixed
  outward into the larger lobe remains. A bulk backflow, with transport of
  and mixing of plasma throughout the lobe would be required, but seems
  less plausible given the lobe morphology.

Considering the merits of these three possibilities, we favour scenario (ii), and suggest a model for the expansion history of the source in which, 75 Myr ago, an outburst swept its way through the X-ray gas, beginning the excavation of the cavities. These cavities grew in stages as the jet power varied over time, with the lobe plasma being refreshed by each new outburst or period of high jet power, the most recent of which is only $\sim$30 Myr old. The overall lobe/cavity inflation time would thus be similar to the X-ray estimate, but the plasma we see in the lobes is a mixture of old and new particles injected by a period of stronger jet activity over the last $\sim$30-40 Myr.

\subsection{Pressure balance and particle content}
\label{sec:press}

The non-thermal energy density in the lobes of an extended radio source cannot be determined by the radio data alone. Only a minimum value can be estimated when we assume the contribution from the magnetic field and the relativistic particles equal \citep{Burbidge58}. The additional information needed to constrain the internal energy density can be obtained either from measurement of the inverse-Compton emission caused by the up-scattering of photons off the relativistic electrons in the lobes, or by assuming pressure equilibrium between the lobes and their environment and measuring the IGM pressure using X-ray data \citep{Longair73}.

We first test our ability to detect inverse-Compton emission from the lobes. Assuming equipartition, \citet{ewan4261} estimated an expected X-ray flux density from up-scattered cosmic microwave background (CMB) photons to be $\sim$2~nJy from each lobe at 1~keV. Using our improved constraints on the properties of the lobe plasma, we repeat this calculation, and find that we expect a flux of $\sim$6.5$\times$10$^{-14}$~erg~s$^{-1}$ in the 0.5-7~keV band ($\sim$4.1~nJy at 1~keV) from each lobe. We then reprocessed the longest \textit{XMM-Newton} dataset available (ObsID 0502120101, $\sim$70~ks after cleaning) using the latest calibration and most recent version of the \textit{XMM} Science Analysis System (\textsc{sas} 13.5). Following the techniques described in O'Sullivan et al., we find no detection of inverse-Compton flux from the lobes, but are able to place 3$\sigma$ upper limits on the 0.5-7~keV flux of $<$6$\times$10$^{-14}$~erg~cm$^{-2}$~s$^{-1}$ ($<$3.8~nJy) for the west lobe, and $<$6.2$\times$10$^{-14}$~erg~cm$^{-2}$~s$^{-1}$ ($<$4.0~nJy) for the east lobe. These limits are marginally below the predicted flux, and so may suggest that the lobe plasma is out of equipartition, with more energy in the magnetic field than in the particle population. However, the uncertainties in the radio fluxes, \textit{XMM} calibration \citep[see, e.g.][\citealt{Tsujimoto11}]{Nevalainen10} and the choice of X-ray spectral model only weakly constrain departures from equipartition in the sense of domination by the electron (and positron) energy density. Deeper X-ray data are needed to provide a lower flux limit and determine whether the lobes are truly in equipartition.

We also estimated the likely optical flux in the inner part of the lobes from up-scattering of infra-red photons emitted by the stellar population. Unfortunately, the expected signal is low, a factor $\sim$10$^{-3}$ below the optical luminosity of the stellar population in the same region. We are therefore unable to constrain the properties of the lobe emission from its inverse-Compton emission, at least with the currently available data.

\begin{figure}
\includegraphics[trim= 1.8cm 6.5cm 0cm 3.7cm, clip=true, width=\columnwidth]{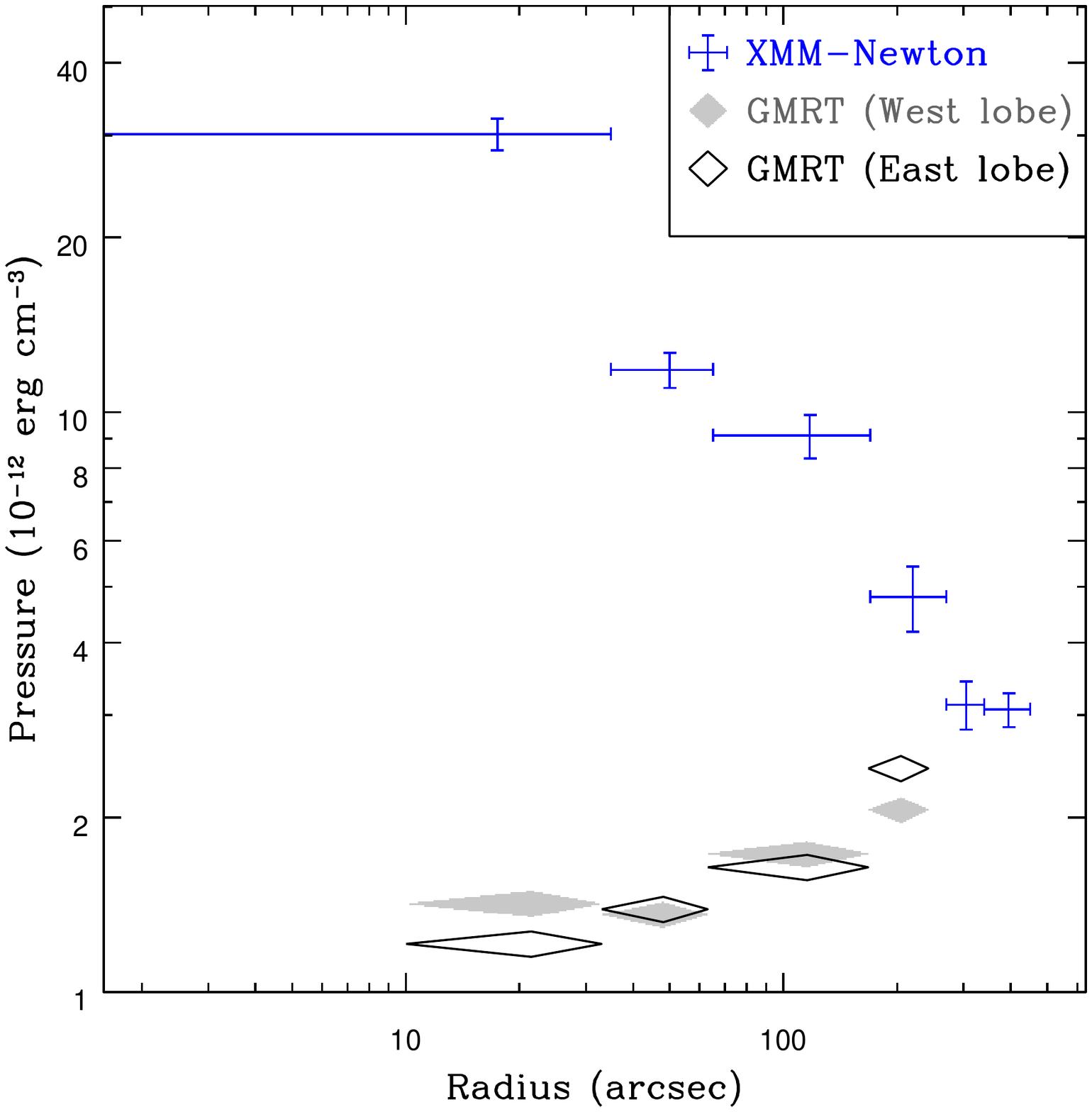}
\caption{\label{fig:press}Deprojected pressure profile of X-ray thermal pressure and radio pressure with radius from the center of the source.}
\end{figure} 

As inverse-Compton emission does not provide a definitive constraint, we compare the apparent lobe pressure with that of the surrounding IGM. Figure~\ref{fig:press} shows the pressure profiles of the thermal gas ($P_{gas}$) along with the profile of the radio pressure derived in this study, as a function of the distance from the source's centre. The X-ray pressure profile is drawn from the XMM-Newton analysis of \citet{ewan4261}. As can be seen from Fig.~\ref{fig:press} the X-ray pressure profile declines with radius as expected, while the pressure profile derived from the radio lobes rises with radius. The radio lobes appear to be underpressured with respect to the surrounding environment in all regions and in both lobes by a factor ranging from $\sim$2 to $\sim$30. We note that the \citet{ewan4261} IGM pressure profile is in excellent agreement with the earlier measurements of \citet{Croston08}, but that Croston et al. found a mean internal pressure for the radio lobes of $\sim$9$\times$10$^{-13}$~erg~cm$^{-3}$, slightly lower than the pressures we measure in the lobes.

A similar pressure imbalance has been found in previous lobe studies of other FR I radio sources, where $P_{gas}$/$P_{radio}$ is usually $>$1, reaching values up to 100, assuming the standard equipartition arguments \citep[e.g.][]{Feretti92, Dunn05, Croston08, Birzan08, Dunn10}. This can only suggest two possible situations for the radio lobes in the case of 3C 270: 1) Deviation from equipartition within the radio lobes with a higher magnetic field energy density than calculated, or 2) the equipartition argument is still valid but either the radio lobes contain a significant amount of non-radiating particles that contribute to the internal pressure or the filling factor $\phi$ of the plasma is less than unity. Since we believe that jets entrain material in FR I galaxies and in this specific galaxy we have reason to think the filling factor might be $\leq1$ at the back of the lobes (volume overestimation including part of the brighter X-ray emitting gas near the cool core), we examine the implications of the lack of pressure balance in the case where equipartition stands.

Assuming that the radio lobes of 3C 270 are in approximate pressure equilibrium with the IGM, we have the ability to constrain the particle content of the lobes from the comparison of the apparent pressures of the relativistic plasma of the radio lobes and that of the thermal plasma of the IGM \citep[e.g.,][]{HardcastleWorrall00,Birzan08}.

In Table~\ref{tab:energy}, the results shown for $(1+k)/\phi$ estimation were calculated by forcing $P_{radio}$ to equal $P_{Xray}$ (note that in Table~\ref{tab:SGfit} we adopted a fixed $(1+k)/\phi=2$). The regions of the radio pressure in this study were chosen so as to match the X-ray annuli used by \cite{ewan4261}. Both lobes show the same trend with the $(1+k)/\phi$ factor rising as we move from the end of the jets (regions 4 and 8) to the back of the lobes near the core (regions 1 and 5). The size of the bins were not exactly equal so the increase of $(1+k)/\phi$ is not linear.  

Assuming that the filling factor $\phi=1$, the additional pressure required
by the rise of $(1+k)/\phi$ in regions 2-4 and 6-8 can be provided
  by non-radiative particles in the relativistic plasma of the lobes.
This non-radiating component could be introduced via the
entrainment and heating of thermal plasma by the radio jets. A probable
source of entrained material is stellar mass loss from stars within the jets \citep{Phinney83,Komissarov94,Hardcastle03,Hardcastle07}. In \citet{ewan4261} the rate of stellar mass-loss from AGB stars into each jet was estimated to be
1.67$\times$$10^{-3}$~M$_\odot$~yr$^{-1}$. By assuming a 37 Myr
  timescale for the AGN outburst, we expect a minimum of $\sim$60000
  $M_\odot$ of stellar material to be entrained by each jet.

\begin{table*}
\begin{center}
\caption{\label{tab:energy}Energy estimates for the particle content of the radio lobes}
\begin{tabular}{lccccc}
\hline
 Region   &     E$_e$       &    $P_{Xray}$           &  V$_{tot}$            &   $(1+k)/\phi$  & Total energy   \\
          & ( 10$^{56}$ erg)&($10^{-12}$dyn cm$^{-2}$)&  ( 10$^{67}$  cm$^{3}$)&             & ( 10$^{56}$ erg )  \\
\hline
West Lobe &              &          &            &               &                      \\
  1       &   1.1    & 30.2  & 5.1  &  42.5  &    14.6        \\
  2       &    2.3   & 11.8  & 11.3  &  17.4  &    11.9      \\ 
  3       &    9.2   &  9.1  & 35.5 &   10.5 &    26.2     \\ 
  4       &   2.3  &  4.8   & 7.5      &      4.7 &  2.1    \\ 
\hline
East Lobe &           &               &             &           &        \\
  5      &  0.6       &   30.2    &   3.1    &    49.8    &   9    \\ 
  6      &  1      &   11.8     &   4.8   &      17    &   5.1 \\
  7      &  7.1    &    9.1   &    28.7    &     11.1  &    21.4    \\ 
  8      &  5      &     4.8   &  13.7      &    3.9   &  3.3    \\ 
\hline
\end{tabular}
\end{center}
\end{table*}

Thermal plasma entrained from stars within the radio jets
  represents a plausible minimum mass of entrained material. Taking the
  limiting assumption that in regions 4 and 8 the only entrained material
  is that from stars in the jets (i.e., that no additional entrainment
  occurs across the lobe/IGM boundary), we can estimate an upper limit on
  the temperature to which this thermal component must be heated in order
  to bring the lobes into thermal balance with their surroundings. We find
  the temperature of the entrained material to be k$_B$T$\la$11 MeV which
corresponds to a plasma with electrons having a Lorentz factor
$\gamma$$\la$20, well into the relativistic regime. Electrons of
$\gamma$$\sim$20 are expected to radiate in the equipartition
magnetic field at $\sim$1 kHz, making them invisible in the observable
radio band. Comparable values have been found by applying similar arguments to Cen A \citep{Wykes13} and 3C~31 \citep{CrostonHardcastle14}.    The very low density of this entrained thermal plasma
component ($n\sim1.4 \times 10^{-7}$ cm$^{-3}$) means that no significant
thermal or non-thermal X-ray emission is expected from it. By
summing up $(P_{Xray}-P_{radio})V$ from each region, where V is the
volume of each region, we estimate that the total energy
(Table~\ref{tab:energy}) required to heat all the entrained material in the
lobe to 11 MeV is $\sim 9\times 10^{57}$ erg.

Alternatively, pressure balance could be produced if the filling factor $\phi<1$. This could occur in several ways, for example if we have overestimated the lobe volume or if significant variations in the density of relativistic particles and magnetic field exist within the lobes.

The most likely location in which $\phi$ may be $<$1 is in regions 1 and 5, which have values of $(1+k)/\phi$ much greater than those of the neighbouring regions. Since regions 1 and 5 of the lobes overlap with the brighter X-ray emission north and south of the cool core, it is possible either that our assumption that the morphology of the lobe can be described as an annulus in this region is oversimplistic, or that the lobe and IGM plasmas are mixed, with filaments or clouds of IGM gas penetrating the volume apparently occupied by the lobe. If we estimate the value of k based on the increase in k from regions 1-3 and 5-7, we find that we would expect $k\sim$24 for region 1 and $k\sim$23 for region 5. This suggests that $\phi$ could be as low as $\sim$0.59 for region 1 and $\sim$0.48 for region 5.

If thermal plasma is present in the lobes, we would expect it to cause Faraday rotation. We can therefore compare our expected density of entrained material with limits derived from the rotation measures for 3C~270 found by \citet{Guidetti11} and \citet{Laing14}. While these measurements include the effects of external Faraday rotation, they do provide an upper bound on the internal rotation measure. Following \citet{Feretti99} we calculate the expected rotation measure from:  

\begin{equation}
RM=812\int_0^L \mathrm{n_{e} B_{z} \, dz}
\label{eqn:RM}
\end{equation}

where RM is the rotation measure in rad m$^{-2}$, n$_{\rm e}$ is the electron number density in cm$^{-3}$, B$_{\rm z}$ is the magnetic field along the line-of-sight in $\rm \mu$G and L is the integration path (the depth of the lobe along the line-of-sight) in kpc. Based on the Guidetti et al. and Laing \& Bridle results, an RM of order 10 rad~m$^{-2}$ seems representative for the system. We find that an electron gas density of $\sim$8.2$\times$10$^{-5}$ cm$^{-3}$ would be required to produce RM=10 rad~m$^{-2}$ purely from internal depolarization. The expected electron density of thermal plasma entrained in the lobes ($\sim$1.4$\times$10$^{-7}$ cm$^{-3}$) is low compared to this limit, and would be expected to produce RM$\sim$0.02 rad~m$^{-2}$, well below the sensitivity of the \citet{Guidetti11} maps. We therefore agree with the conclusion drawn by Guidetti et al., that the RM structures they observe are probably caused by foreground depolarization.

We can also test whether regions 1 and 5, which we earlier suggested have filling factors less than unity, could contain a mix of thermal and relativistic plasmas, or whether it is more like that our adopted geometry leads us to overestimate their volume. Using equation~\ref{eqn:RM} and the IGM electron density derived from the X-rays in \citet{ewan4261} we find that we expect an internal RM of $\sim500$ rad~m$^{-2}$ for both regions. While \citet{Guidetti11} do find evidence of enhanced depolarization at the inner edges of the lobes, our predicted value is a factor $\sim$20 higher than they observe. This indicates that the relativistic and thermal plasmas along these lines of sight are probably not well mixed. This could suggest that the geometry we have
assumed for the lobes close to the core is too simple, for example if filaments or sheets of thermal plasma penetrate the lobe while remaining
magnetically separate. Alternatively, the internal properties of the lobe plasma may differ in these regions from those seen through the rest of the lobes.

In \citet{ewan4261} the total enthalpy, 4PV - the energy required to expand
the lobe to its current size against the pressure of the IGM, plus the
energy stored in the magnetic field and particles in the lobe - was
estimated to be $\sim2.4\times10^{58}$ erg. Our energy estimate of $\sim
9\times 10^{57}$~erg required to heat any entrained material is
about 40\% of the enthalpy estimated in \citet{ewan4261}.
This is a factor of 1.8 greater than estimated by O'Sullivan et
  al., mainly owing to our more accurate treatment of lobe volume. The
estimate changes little if the lower filling factors in regions~1 and~5 are
adopted. 

In summary, we have examined the constraints which can be placed on the particle populations in the radio lobes from the available radio and X-ray measurements. We find that only a relatively small mass of entrained gas is needed to bring the lobes into pressure equilibrium with their surroundings, that this is probably consistent with entrainment of stellar material lost from stars within the jets, and that such material would have a low density and therefore would not have detectable effects on the polarization.


\section{Conclusions}
\label{sec:conc}

In this paper, we have presented a detailed spectral analysis of the FR I radio source 3C 270, emanating from the nearby group-central elliptical galaxy NGC~4261. We have analysed 240 MHz and 610 MHz data from the GMRT, and data from the VLA at 1.55 GHz and 4.86 GHz.

Combining the analysed GMRT and VLA data with data flux density measurements from the literature between 22~MHz and 33~GHz, we used a Jaffe \& Perola (JP) model to determine the asymptotic low-frequency spectral index of 3C~270 to be $\alpha_{inj}=0.53_{-0.02}^{+0.01}$. From the GMRT 240 MHz and VLA 1.55 GHz and 4.8 GHz, we produced spectral index maps of the source which suggest a spectral index of $\alpha\sim$0.16 for the unresolved and probably self-absorbed core, a constant spectral index along the jets of $\alpha\sim$0.5 and a gradual steepening from the tip of the jets through to the back of the lobes of $\alpha\sim$0.8$-$1. 

Assuming equipartition of energy between the magnetic field and the
relativistic particles, and adopting a minimum Lorentz factor of
$\gamma_{\rm min}$=100 for the radiating electrons, we estimate the
radiative age of 3C~270. We modelled the observed 1.55-4.86~GHz
  spectral index trend in the lobes, fitting a Jaffe \& Perola (JP) model.
  We find a radiative age of $t_{\rm rad}\sim29$ and 37 Myr for the west
  and east lobes respectively. This is about half the dynamical age of
  $<$75 Myr found previously using X-ray data \citep{ewan4261}. We
  considered three possible scenarios for this discrepancy. The least
  likely is that the source underwent an early period of supersonic
  expansion. While the shocks from such an expansion might now be outside
  the field of view of the available X-ray observations, they would have
  very significantly heated the IGM, implying unphysically low pre-shock
  temperatures.

A second alternative is that a backflow within the lobes may be
  transporting young, energetic electrons from the jet into regions of
  older plasma, leading us to underestimate the age. This cannot be ruled
  out, but previous high resolution observations of the base of the jets do
  at least argue that no backflow is observed on small scales. Finally, the
  third possibility, which we favour, is the observed source is the product
  of multiple AGN outburst or large change in jet power, with the radiative
  age corresponding to the latest outburst or period of high jet power.
  \citet{Worrall10} noted structures in the X-ray jet that could be
  indicate variations in power on 10$^4$~yr timescales, and larger scale
  variations in jet brightness are observed in the radio. A renewed
  outburst would not only inject a fresh population of young electrons into
  the lobes, but would likely drive shocks into the only population,
  causing reacceleration. The end result would be a source whose lobe
  expansion time might be similar to the X-ray age estimate while the
  radiative age represents the onset of more recent activity.

Assuming that the pressures in the radio lobes equal that of the IGM we calculate that the ratio of energies of non-radiating to radiating particles, k, ranges from $\sim4-24$ from the tip of the jets to the lobes close to the core (Table~\ref{tab:energy}, Fig.~\ref{fig:vol}). This suggests that gas entrainment was more effective at earlier times, when the jets were shorter. Assuming, as a limiting case, that currently only material lost from stars in the jets is entrained, we estimate that the temperature of the entrained material would need to be 11~MeV if it is to bring the lobes into pressure balance with the IGM. This corresponds to a total energy of $\sim10^{58}$ erg required to heat the entrained material in the lobe, $\sim$40\% of the total enthalpy of the lobes estimated in \citet{ewan4261}.

Based on our estimates of the density of the entrained material, we predict the likely internal rotation measure arising from this component, and compare it to RM measurements from the literature. The density of $\sim1.4 \times 10^{-7}$ cm$^{-3}$  indicates that the contribution of the entrained material to the Faraday depolarization is undetectable. In regions 1 and 5 where the thermal X-ray emitting plasma co-exists with the relativistic plasma, we would expect a RM of $\sim500$~rad~m$^{-2}$ for both regions if mixing had taken place. This is ruled out by the observations, implying that little mixing between the thermal and relativistic plasmas has occurred.

\section*{Acknowledgments}
K. Kolokythas is supported by the University of Birmingham. E. O'Sullivan acknowledges support for this work from the National Aeronautics and Space Administration through Chandra Award Number AR3-14014X issued by the Chandra X-ray Observatory Center, which is operated by the Smithsonian Astrophysical Observatory for and on behalf of NASA under contract NAS8-03060, and through the Astrophysical Data Analysis programme, award NNX13AE71G. S. Giacintucci acknowledges the support of NASA through Einstein Postdoctoral Fellowship PF0-110071 awarded by the {\em Chandra} X-ray Center (CXC), which is operated by the Smithsonian Astrophysical Observatory (SAO). GMRT is run by the National Centre for Radio Astrophysics of the Tata Institute of Fundamental Research. Some of this research was supported by the EU/FP7 Marie Curie award of the IRSES grant CAFEGROUPS (247653). We thank T. Ponman and Andreas Zezas for useful discussions of NGC 4261 and formation history. We thank the anonymous referee for his help in improving the paper.


 \label{lastpage}

 \end{document}